\documentclass[12pt]{article}
\usepackage{float}
\usepackage{graphicx}
\graphicspath{ {figs/} }
\usepackage{longtable}
\usepackage{color}
\usepackage{amsmath,bbm,bm,mathtools}
\usepackage[english]{babel}
\usepackage{dsfont}
\usepackage[OT1]{fontenc}
\usepackage[utf8]{inputenc}
\usepackage{subcaption}
\usepackage{refcount}
\usepackage{lmodern}
\usepackage{color}
\usepackage{booktabs}
\usepackage{natbib}
\usepackage{multirow}
\usepackage{setspace}

\usepackage{hyperref}
\usepackage{nameref,zref-xr}
\zxrsetup{toltxlabel}
\usepackage{url}

\usepackage{mathtools}
\usepackage{amsfonts}
\usepackage{amssymb}
\usepackage{bbm}
\usepackage{amsthm}
\usepackage{amsxtra}
\usepackage{amsmath}
\usepackage{commath}
\usepackage{times}
\usepackage{bm}

\usepackage{cleveref}

\makeatletter%
\@ifclassloaded{bio}{%
   \usepackage[nofiglist,nomarkers,figuresonly]{endfloat}

   \newcommand{\titlepaper}{A generalization of moderated statistics to data
      adaptive semiparametric estimation in high-dimensional biology}

   \newcommand{\authorlist}{
     NIMA S.~HEJAZI$^\ast$, PHILIPPE BOILEAU, MARK J.~VAN DER LAAN,
     ALAN E.~HUBBARD
     \\[4pt]
     \textit{
        Division of Biostatistics, Department of Population Health
        Sciences, Weill Cornell Medicine\\[0pt]
        402 E.~67\textsuperscript{th} Street, New York, NY 10065
     }
     \\[6pt]
     \textit{
        Division of Biostatistics, School of Public
        Health, University of California, Berkeley\\[0pt]
        2121 Berkeley Way, Berkeley, CA 94720
     }
     \\[2pt]
     {\texttt{nhejazi@berkeley.edu}}
   }

}{%
  \usepackage{arxiv}
  \usepackage{standalone}
  \usepackage[inline,shortlabels]{enumitem}
  \usepackage{amsfonts,amsthm}
  \usepackage{booktabs}
  \bibliographystyle{biometrika}

   \newcommand{\titlepaper}{A generalization of moderated statistics to data
      adaptive semiparametric estimation in high-dimensional biology}

  \newcommand{\authorlist}{
       Nima S.~Hejazi \\
       Division of Biostatistics, \\
       Department of Population Health Sciences, \\
       Weill Cornell Medicine \\
       \texttt{nhejazi@berkeley.edu} \\
       \And
       Philippe Boileau \\
       Division of Biostatistics, \\
       School of Public Health, and \\
       Center for Computational Biology, \\
       University of California, Berkeley \\
       \texttt{pboileau@berkeley.edu} \\
       \And
       Mark J.~{van der Laan} \\
       Division of Biostatistics, \\
       School of Public Health, and \\
       Department of Statistics, and \\
       Center for Computational Biology, \\
       University of California, Berkeley \\
       \texttt{laan@berkeley.edu} \\
       \And
       Alan E.~Hubbard \\
       Division of Biostatistics, \\
       School of Public Health, and \\
       Center for Computational Biology, \\
       University of California, Berkeley \\
       \texttt{hubbard@berkeley.edu} \\
  }
}
\makeatother%

\newcommand\prob{{\mathbb{P}}}
\newcommand\var{{\mathbb{V}}}
\newcommand{\E}{\mathbb{E}}

\newcommand{\M}{\mathcal{M}}

\newcommand{\paperabstract}{
  The widespread availability of high-dimensional biological data has made the
  simultaneous screening of many biological characteristics a central problem
  in computational biology and allied sciences. While the dimensionality of
  such datasets continues to grow, so too does the complexity of biomarker
  identification from exposure patterns in health studies measuring baseline
  confounders; moreover, doing so while avoiding model misspecification remains
  an issue only partially addressed. Efficient estimators capable of
  incorporating flexible, data adaptive regression techniques in estimating
  relevant components of the data-generating distribution provide an avenue for
  avoiding model misspecification; however, in the context of high-dimensional
  problems that require the simultaneous estimation of numerous parameters,
  standard variance estimators have proven unstable, resulting in unreliable
  Type-I error control even under standard multiple testing corrections. We
  present a general approach for applying empirical Bayes shrinkage to variance
  estimators of a family of efficient, asymptotically linear estimators of
  population intervention causal effects arising from comparing counterfactual
  contrasts of an exposure variable. Our generalization of shrinkage-based
  variance estimators increases inferential stability in high-dimensional
  settings, facilitating the application of these estimators for deriving
  nonparametric variable importance measures in high-dimensional biological
  datasets with modest sample sizes. The result is a data adaptive approach for
  robustly uncovering stable causal associations in high-dimensional data in
  studies with limited samples. Our generalized variance estimator is evaluated
  against alternative variance estimators in numerical experiments, and an open
  source \texttt{R} package for the Bioconductor project, \texttt{biotmle}, is
  introduced. Identification of biomarkers with the proposed methodology is
  demonstrated in an analysis of high-dimensional DNA methylation data from an
  observational study on the epigenetic effects of tobacco smoking.
}

\pdfoutput=1
\zexternaldocument*[supp:]{sm}

\title{\titlepaper}
\date{\today}
\author{\authorlist}

\begin{document}
\maketitle

\begin{abstract}
  \paperabstract
\end{abstract}

\section{Introduction}\label{intro}

High-dimensional biomarker data is now routinely collected in observational
studies and randomized trials in the biomedical and health sciences. The
statistical analysis of such data often relies on parametric modeling efforts
that allow covariate adjustment to obtain inference in samples that are small or
moderately sized relative to biomarker dimensionality. By treating each
biomarker as an independent outcome, standard differential expression analyses
fit biomarker-specific linear models while adjusting for potential baseline
confounders in the model's postulated form, capturing the effect of a common
exposure on each biomarker when the parametric form is \textit{correctly
specified}. While the underlying asymptotic theory of linear models is robust,
these techniques have been adapted for use in small-sample settings through
variance moderation (or shrinkage) approaches, which stabilize inference on the
relevant parameter of the linear model. The moderated t-statistic, the most
popular among such approaches, was first formulated through a hierarchical model
based on empirical Bayes shrinkage of the standard error estimates of the target
parameter~\citep{smyth2004linear}; its corresponding implementation in the
\texttt{limma} software package for the \texttt{R} programming
language~\citep{R} has been heavily utilized in studies using microarray and
next-generation sequencing data~\citep{smyth2005limma, law2014voom}. We
generalize this variance moderation strategy to a broad class of efficient,
asymptotically linear estimators, increasing their robustness in settings with
a limited number of independent units.

Given a high-dimensional biological dataset, a standard differential expression
analysis pipeline proceeds by fitting a common-form linear model individually to
each of the many candidate biomarkers, using an exposure as the primary
independent variable and adjusting for potential confounders of the
exposure--outcome relationship by the addition of main terms to the parametric
functional form. To stabilize inference, the moderated t-statistic may be used
to shrink variance estimates towards a common value across the candidate
biomarkers~\citep{smyth2004linear}, alongside multiplicity corrections to adjust
for testing many hypotheses~\citep{dudoit2008multiple}. Within this framework,
the estimated coefficient of the exposure would be taken as an estimate of the
scientific quantity of interest --- that is, the causal effect of the exposure
on the expression of candidate biomarkers. While it is common practice, such an
approach is rarely rooted in available scientific knowledge, requiring unfounded
assumptions (e.g., postulating an exact linear form) to be introduced by the
analyst. A common pitfall in standard practice is misspecification of this
parametric form, which leads to the target estimand being misaligned with the
motivating scientific question. Only recently have tools from modern causal
inference~\citep[e.g.,][]{pearl2000causality} been recognized as offering
rigorous solutions to such issues in observational biomarker
studies~\citep[e.g.,][]{reifeis2020assessing, reifeis2020causal}.

A rich literature has developed around the construction of techniques that
eschew parametric forms, relying instead on developments in non/semi-parametric
inference and machine learning~\citep{bembom2009biomarker,vdl2011targeted} to
avoid the pitfalls of model misspecification. By targeting nonparametric
estimands and performing model fitting via automated, data adaptive regression
techniques~\citep{vdl2006targeted,vdl2007super}, such non/semi-parametric
procedures exhibit a robustness that is generally desirable. Unfortunately,
a common limitation in their application is the mutual incompatibility of
machine learning-based strategies, convergence rates required for asymptotic
statistical inference, and the limited sample sizes available in biomarker
studies. Since non/semi-parametric estimation approaches generally converge at
much larger sample sizes than their parametric
counterparts~\citep{vdl2011targeted}, these approaches can suffer from variance
estimation instability in even modestly sized studies and thus stand to benefit
from variance moderation at such sample sizes.

Our principal contribution is an adaptation of an empirical Bayes shrinkage
estimator, or variance moderation, to derive stabilized inference for data
adaptive estimators of nonparametric estimands. Specifically, through the
comparison of four non/semi-parametric variance estimation strategies, we
demonstrate that a generalized variance shrinkage approach can improve the
stability of efficient, data adaptive estimation procedures in small and
modestly sized biomarker studies. We introduce a modified reference distribution
for hypothesis testing with moderated test statistics, further strengthening the
Type-I error control of our biomarker identification strategy. We emphasize that
our proposal need not be a competitor to other marginal variance stabilization
strategies formulated for non/semi-parametric efficient estimators; rather, it
may be coupled with such methods to further stabilize the resultant variance
estimates.

Our approach may be applied directly to a wide variety of parameters commonly of
interest, as long as an \textit{asymptotically linear estimator} of the target
parameter exists. Such estimators are characterized by their asymptotic
difference from the target parameter admitting a representation as the sum of
independent and identically distributed random variables (i.e., the estimator's
influence function). Asymptotically linear estimators have been formulated for
both parameteric estimands and nonparametric estimands defined in causal
models~\citep{vdl2011targeted}. While our variance moderation approach may be
applied in a vast array of problems, its advantages are particularly noteworthy
in high-dimensional settings, when the sampling distributions of complex,
non/semi-parametric efficient estimators are often erratic and prone to yielding
high false positive rates.

The remainder of the present manuscript is organized as follows.
Section~\ref{background} briefly introduces elements of both classical variance
moderation non/semi-parametric theory and locally efficient estimation with
asymptotically linear estimators in the nonparametric model.
Section~\ref{methodology} details the proposed approach, including an
illustration of generalizing variance shrinkage to a non/semi-parametric
efficient, doubly robust estimator of the average treatment effect, alongside
a robustified moderated test statistic. The results of interrogating the
proposed technique in simulation experiments are then presented in
Section~\ref{simulation}, evaluating performance against a popular
variance-moderated linear modeling strategy and non/semi-parametric efficient
estimators without variance moderation. In Section~\ref{analysis}, we
demonstrate our approach by applying our variance-moderated doubly robust
estimation procedure to evaluate evidence from an observational
study~\citep{su2016distinct} on the epigenetic alterations to DNA methylation
biomarkers caused by tobacco smoking. Section~\ref{discussion} concludes by
summarizing our findings and by identifying avenues for future investigation.

\section{Preliminaries and Background}\label{background}

\subsection{Data, notation, and statistical model}\label{prelim}

We consider data generated by typical cohort sampling, where the data on a
single observational unit is denoted by the random variable $O = (W, A, Y)$,
where $W \in \mathcal{W}$ is a vector of baseline covariates, $A \in
\mathcal{A}$ is a binary exposure, and $Y = (Y_b, b = 1, \ldots, B) \in
\mathcal{Y}$ is a vector of outcomes, like candidate biomarker measurements. We
assume access to $n$ independent copies of $O$, using $P_0$ to denote the
distribution of $O$. Further, we assume a nonparametric statistical model $P_0
\in \mathcal{M}$ composed of all distributions subject to some dominating
measure, thereby placing no restrictions on the form of $P_0$.
Let $q_{0, Y}$ denote the conditional density of $Y$ given $(A, W)$ with
respect to dominating measure $\mu$; $g_{0, A} \coloneqq \prob(A = 1 \mid W)$,
the conditional probability of $A$ given $W$; and $q_{0, W}$ the density of $W$
with respect to dominating measure $\nu$. We use $p_0$ to denote the density of
$O$ with respect to the product measure. Evaluated on a typical observation $o$,
this density $p_0$ is
$p_0(o) = q_{0,Y}(y \mid A = a, W = w) g_{0,A}(a \mid W = w) q_{0,W}(w)$.

A nonparametric structural equation model (NPSEM) allows for counterfactual
quantities of interest to be described by hypothetical interventions on the
data-generating mechanism of $O$~\citep{pearl2000causality}. We assume an NPSEM
composed of the following system of equations:
$W = f_W(U_W), A = f_A(W, U_A), Y = f_Y(A, W, U_Y)$,
where $f_W$, $f_A$, and $f_Y$ are deterministic functions, and $U_W$, $U_A$,
and $U_Y$ are exogenous random variables. The NPSEM provides a parameterization
of $p_0$ in terms of the distribution of the endogenous and exogenous random
variables modeled by the system of structural equations, implying a model for
the distribution of counterfactual random variables generated by specific
interventions on the data-generating process. For simplicity, we consider
a \textit{static intervention}, defined by replacing $f_A$ with a value $a \in
\mathcal{A}$, the support of $A$. Such an intervention generates
a counterfactual random variable $Y(a) = (Y_b^{a}, b: 1, \ldots B)$, defined as
the values the $B$ candidate biomarker outcomes would have taken if the exposure
$A$ had been set to level $a \in \mathcal{A}$, possibly contrary to fact.

Although our proposal applies to any asymptotically linear estimator, we
will focus on efficient estimators of the average treatment effect (ATE) in the
sequel, as the ATE is a canonical, well-studied causal parameter. The ATE
$\psi_b$ may be defined as the expected population-level difference,
between the counterfactual expression of a given candidate
biomarker when the static intervention is imposed and its counterfactual
expression when the intervention is withheld, marginalizing over all strata of
$W$. That is, the ATE may be expressed  as $\psi_b = \E_0[Y_b(1) -
Y_b(0)]$~\citep{pearl2000causality}, where $Y_b(1)$ is the potential outcome of
candidate biomarker $b$ when the static intervention is applied and $Y_b(0)$ the
potential outcome in the absence of the exposure. Throughout, we opt for
nonparametric statistical estimands rooted in causal inference on account of
their close alignment with scientifically informative quantities.

\subsection{Asymptotic linearity and influence functions}\label{eif}

As our proposal generalizes the approach of the moderated
t-statistic~\citep{smyth2004linear}, we first examine how a typical data
analysis may be conducted with the outlined data structure. As the same strategy
is applied to obtain marginal estimates of biomarker importance for every
biomarker $b = 1, \ldots, B$, we will focus on only a single biomarker $Y_b$,
suppressing dependence on the index $b$ in the sequel.

For a binary exposure $A$ and single binary baseline covariate $W$, assume the
relationship of the exposure with the outcome is characterized by a working
linear model $m_{\beta}$, i.e., the projection $\E[Y \mid A, W] = \beta_0
+ \beta_1 A + \beta_2 W$. The scientific quantity of interest --- the effect of
exposure on the expression of the biomarker $Y$, controlling for the effect of
the baseline covariate $W$ --- is captured by the model parameter $\beta_1$.
Since $A \in \{0, 1\}$, the parameter of interest $\beta_1$ is a difference in
conditional means of the exposure groups. The estimator $\hat{\beta}_1$ of
$\beta_1$ is characterized as \textit{asymptotically linear} by the fact that it
may be represented in terms of a mean-zero function:
\begin{equation*}\label{simple_if}
  \sqrt{n}(\hat{\beta}_1 - \beta_1) = \frac{1}{\sqrt{n}}\sum_{i = 1}^n D(O_i) +
    o_p(1),
\end{equation*}
where $D(O_i) = C^{-1}(A_i, W_i) (Y_i - m_{\beta}(A_i, W_i))$ is the influence
function of $\beta_1$ and $C = \E[(A_i, W_i)(A_i, W_i)^{T}]$. The influence
function characterizes the asymptotic difference between the estimator
$\hat{\beta}_1$ and the parameter $\beta_1$ as such,
\begin{equation}\label{simple_if_dist}
  \sqrt{n}(\hat{\beta}_1 - \beta_1) \xrightarrow[D]{} N(0, \sigma^2(D)),
\end{equation}
where the limit distribution is mean-zero normal with variance matching that
of the influence function. A $(1- \alpha)$ Wald-style confidence interval for
$\beta_1$ may be constructed straightforwardly as $\hat{\beta}_1 \pm \{Z_{(1 -
\alpha/2)} \hat{\sigma}(D)\} / \sqrt{n}$, where $\hat{\sigma}^2(D)$ is the
empirical variance of the estimated influence function.

Importantly, while an estimator may admit non-unique representations in terms of
several influence functions in constrained statistical models, an asymptotically
linear estimator has only a single unique influence function in the
nonparametric model $\M$, often called the \textit{efficient influence function}
of the estimator. The form of the efficient influence function is a key
ingredient in the construction of regular asymptotically linear estimators
capable of achieving the non/semi-parametric efficiency
bound~\citep{bickel1993efficient, vdl2011targeted}.

\subsection{Empirical Bayes variance moderation}\label{limma}

Variance moderation has been established as a promising and useful tool for
stabilizing test statistics. The general methodology consists in the application
of a shrinkage estimator to the individual variance estimates across a large
number of (related) hypothesis tests. The moderated t- and
F-statistics~\citep{smyth2004linear} are perhaps the most commonly used examples
of variance moderation approaches in differential expression analysis.
Considering the same linear modeling approach previously formulated, a typical
differential expression analysis would fit $B$ linear models $\hat{Y}_b
= \hat{\beta}_{0,b} + \hat{\beta}_{1,b} A + \hat{\beta}_{2,b} W$, using
a standard or moderated test statistic to assess the effect of $A$ on each of
the $B$ biomarkers marginally. The moderated t-statistic~\citep{smyth2004linear}
takes the form
\begin{equation}\label{moderated_t}
  \tilde{t}_b = \frac{\hat{\beta}_{1,b}}{\tilde{\sigma}_b}
  \quad \textrm{where} \quad
  \tilde{\sigma}_b^2 = \frac{d_0 \hat{\sigma}_0^2 + d_b
    \hat{\sigma}_b^2}{d_0 + d_b},
\end{equation}
in which $d_b$ and $d_0$ are the degrees of freedom for the
$b$\textsuperscript{th} biomarker and the remaining ($B - 1$) biomarkers,
respectively, and $\hat{\sigma}_b$ is the standard deviation for the
$b$\textsuperscript{th} biomarker while $\hat{\sigma}_0$ is the standard
deviation across all other biomarkers.

The resultant test statistic has much the same interpretation as an ordinary
t-statistic, though its standard error is now shrunken towards a common value
(i.e., moderated) across all biomarkers based on a hierarchical Bayesian
model~\citep{smyth2004linear}. The process of generating p-values for the
moderated t-statistic is analogous to that of the ordinary t-statistic, with the
only difference being that the degrees of freedom may be inflated to account for
the increased robustness of moderated test statistics~\citep{smyth2004linear}.
The approach was introduced in the \texttt{limma} \texttt{R} package, available
via the Bioconductor project~\citep{smyth2005limma, gentleman2004bioconductor};
it remains extremely popular for biomarker identification and differential
expression analysis across many domains today.

\subsection{Targeted variable importance measures}\label{targetparam}

In the high-dimensional settings common in biomarker discovery studies, the
tools of causal inference and non/semi-parametric theory may be leveraged to
develop efficient estimators of the effect of an exposure on an outcome while
flexibly controlling for unwanted effects attributable to potential confounders.
Commonly, variable importance analyses seek to derive rankings of the relative
importance of candidate biomarkers based on their independent associations with
another variable of interest, such as exposure to an environmental toxin or
disease status~\citep{bembom2009biomarker, tuglus2011targeted, vdl2011targeted}.

To proceed, we define the target parameter as a variable importance measure
based on the statistical functional corresponding, under standard identification
assumptions~\citep{pearl2000causality}, to a causal parameter. We consider
observing $O_1, \ldots, O_n$, i.e., $n$ i.i.d.~copies of the random variable
$O$, the observed data on a single unit. The target parameter $\Psi(P_{0})$ is
defined as a function $\Psi$ mapping the true probability distribution $P_0 \in
\mathcal{M}$ of $O$ into a target feature of interest. Letting $P_n$ denote the
empirical distribution of the observed data, an estimate of the target parameter
$\psi_n$ may be viewed as a mapping from $\mathcal{M}$ to the parameter space
$\Psi$~\citep{vdl2011targeted}. By casting the target parameter as a feature of
the (unobserved) true probability distribution $P_0$, this definition allows
a much richer class of target features of interest than the more restrictive
view of considering only coefficients in possibly misspecified parametric forms.
While we focus on cases where $O_1, \ldots, O_n$ are i.i.d., we note that the
proposed methodology generalizes, with only minor modification, to cases in
which the observed units are clustered, such as when repeated samples on the
same biological unit (i.e., technical replicates) are available.

Prior proposals~\citep[e.g.,][]{bembom2009biomarker} defined a variable
importance measure based on the ATE as
\begin{equation}\label{target_param}
  \psi_b \equiv \Psi_b(P_0) \coloneqq \E_0[\E_0(Y_b \mid A = 1, W) -
    \E_0(Y_b \mid A = 0, W)],
\end{equation}
for a single biomarker $b$.
The target parameter of Equation~\eqref{target_param} is the statistical
functional corresponding to the ATE under identification assumptions standard in
causal inference, including no unmeasured confounding and
positivity~\citep{pearl2000causality}. When these assumptions hold, $\psi_b$ may
be interpreted as the causal difference in the mean expression of the biomarker
under two counterfactual contrasts defined by static interventions on the binary
exposure $A$~\citep{pearl2000causality}; however, even when these assumptions
are unsatisfied, the statistical target parameter is endowed with
a straightforward interpretation: it is the adjusted mean difference in
candidate biomarker expression across exposure contrasts, marginalizing over
strata of potential baseline confounders~\citep{vdl2011targeted}. Finally, if
the true outcome model is, in fact, captured by a linear form (e.g., $\E(Y_b
\mid A, W) = \beta_0 + \beta_1 A + \beta_2 W$), then the ATE corresponds exactly
with $\beta_1$; thus, the estimand conveniently reduces to $\beta_1$ if the
parametric form is correct.

Efficient estimators may be constructed as solutions to the efficient influence
function (EIF) estimating equation $D(O_i)$. For the biomarker-specific ATE
$\psi_b$, the form of the EIF is
\begin{equation}\label{eqn:eif_ate}
  D_b (O_i) = \left[ \frac{2A_i - 1}{g_0(A_i \mid W_i)} \right] (Y_{b, i} -
  \overline{Q}_{0,b}(A_i, W_i)) + \overline{Q}_{0,b}(1, W_i) -
  \overline{Q}_{0,b}(0, W_i) - \psi_b.
\end{equation}
In Equation~\eqref{eqn:eif_ate}, $D_b(O_i)$ is the EIF evaluated at an observed
data unit $O_i$, $\overline{Q}_{0, b}(A,W) = \E(Y_b \mid A, W)$ is the outcome
regression (with corresponding estimator $\overline{Q}_{n,b}$) evaluated at
values of the intervention $A \in \{0,1\}$, and $g_0(A \mid W) = \prob(A
= 1 \mid W)$ is the propensity score (with corresponding estimator $g_n$).
Classical estimators of the ATE (e.g., inverse probability weighting) require
access to either the propensity score or outcome regression, while
non/semi-parametric efficient estimators based on the EIF require estimation of
both nuisance parameters.

\subsection{Data adaptive efficient estimation}\label{tmle}

Several approaches exist for constructing efficient estimators based on the EIF.
Among these, two popular frameworks incorporate data
adaptive regression: one-step estimation~\citep{bickel1993efficient} and
targeted minimum loss (TML) estimation~\citep{vdl2006targeted,vdl2011targeted}.
Both strategies begin by first estimating the nuisance parameters $(g_0,
\overline{Q}_{0,b})$, proceeding to then employ distinct bias-correcting
procedures in their second stages. The resultant estimators, regardless of the
framework used, are consistent when either of the nuisance parameters is
correctly estimated (i.e., doubly robust) and asymptotically achieve the
non/semi-parametric efficiency bound (i.e., the minimum possible variance among
all regular asymptotically linear estimators) when both are accurately
estimated.

\subsubsection{Constructing initial estimators:}

Both classes of efficient estimators accommodate flexible, data adaptive
regression (i.e., machine learning) for the construction of initial estimates of
the nuisance parameters $(g_0, \overline{Q}_{0,b})$, sharply curbing the risk
for model misspecification. Considering the vast and constantly growing array of
machine learning algorithms in circulation, it can be challenging to select
a single algorithm or family of learning algorithms for optimal estimation of
$(g_n, \overline{Q}_{n,b})$. Two strategies for addressing this challenge
include model selection through a combination of cross-validation and loss-based
estimation~\citep{vdl2004asymptotic, dudoit2005asymptotics} and model
ensembling~\citep[e.g.,][]{breiman1996stacked}. The Super Learner
algorithm~\citep{vdl2007super} unifies these strategies by leveraging the
asymptotic optimality of cross-validated loss-based
estimation~\citep{dudoit2005asymptotics} to either select a single algorithm or
produce a weighted ensemble from a user-specified candidate library via
empirical risk minimization of an appropriate loss function. The result is an
asymptotically optimal procedure for estimation of the nuisance parameters
$(g_n, \overline{Q}_{n,b})$, more aptly capturing their potentially complex
functional forms.
A modern implementation of the Super Learner algorithm is available in the
\texttt{sl3}~\citep{coyle2021sl3} \texttt{R} package.

\subsubsection{Efficient estimation:}

In one-step estimation, the empirical mean of the estimated EIF is added to the
initial plug-in estimator, i.e.,
$\psi_{n,b}^{+} = n^{-1} \sum_{i = 1}^{n}
  \left[\overline{Q}_{n,b}(1, W_i) - \overline{Q}_{n,b}(0, W_i)\right] +
  D_{n,b}(O_i)$,
where
$D_{n,b}(O_i) = [(2A_i - 1) / g_n(A_i \mid W_i)] (Y_{b,i}
- \overline{Q}_{n,b}(A_i, W_i)) + \overline{Q}_{n,b}(1, W_i)
- \overline{Q}_{n,b}(0, W_i) - \psi_{n,b}$
is the EIF evaluated at the initial nuisance parameter estimates $(g_n,
\overline{Q}_{n,b})$. TML estimation takes the alternative approach of tilting
the nuisance parameters of the plug-in estimator to solve critical score
equations based on the form of the EIF. The TML estimator is
$\psi_{n,b}^{\star} = n^{-1} \sum_{i = 1}^{n}
  \overline{Q}_{n,b}^{\star}(1, W_i) - \overline{Q}_{n,b}^{\star}(0, W_i),$
where $\overline{Q}_{n,b}^{\star}$ is a tilted version of the initial estimate
$\overline{Q}_{n,b}$
of the outcome regression. The tilting procedure perturbs the initial estimate
$\overline{Q}_{n,b}$ via a one-dimensional parametric fluctuation model, i.e.,
$\text{logit}(\overline{Q}^{\star}_{n,b}(A, W)) =
  \text{logit}(\overline{Q}_{n,b}(A, W)) + \epsilon_n h(A, W)$,
where the initial estimate $\overline{Q}_{n,b}(A, W)$ is treated as an offset
(i.e., coefficient fixed to $1$) and $\epsilon_n$ is the coefficient of the
auxiliary covariate $h(A,W) = (2A - 1) / g_n(A \mid W)$, which incorporates
inverse probability weights based on $g_n(A \mid W)$. When $g_n$ takes extreme
values (close to the boundaries of the unit interval), the fluctuation model may
instead include $h(A,W)$ as a weight, which could improve estimation stability.
The TML estimator $\psi_{n,b}^{\star}$ of $\psi_b$ is derived using the tilted
estimates $\overline{Q}^{\star}_{n,b}$. Owing to their bias-correcting steps,
both the one-step estimator $\psi_{n,b}^{+}$ and the TML estimator
$\psi_{n,b}^{\star}$ have asymptotically normal limit distributions, allowing
for inference based on Wald-style confidence intervals and hypothesis tests.


\subsubsection{Variance estimation based on the efficient influence function:}

As implied by Equation~\eqref{simple_if_dist}, the standard variance estimator
for asymptotically linear estimators is $\mathbb{V}(D_b(O)) / n$. The empirical
variance of the EIF evaluated at initial estimates of the nuisance parameters,
i.e., $\sigma_{n,b}^2 = \var D_{n,b} = n^{-1} \sum_{i=1}^n D^2_{n,b}(O_i)$,
is a valid, occasionally conservative variance estimator for both the one-step
and TML estimators. Thus, asymptotically correct confidence intervals and
hypothesis tests for these efficient estimators may use this variance estimator.
A popular alternative approach instead uses the empirical variance estimator
based on the cross-validated EIF, which addresses issues of overfitting of
nuisance function estimates.
Though this approach improves marginal variance estimates $\sigma_{n,b}^2$, it
fails to take advantage of the benefits that pooled variance estimation may
confer in settings with many outcomes.

Since we advocate for the use of data adaptive regression techniques for
nuisance parameter estimation, we wish to draw particular attention to the
cross-validated variance estimator based on the EIF. Analogous to the
full-sample variance estimator, this estimator is based on the empirical
variance of the EIF evaluated at cross-validated initial estimates of the
nuisance functions. To define such an estimator, denote by ${\cal V}_1, \ldots,
{\cal V}_K$ a random partition of the index set $\{1, \ldots, n\}$ into $K$
validation sets of roughly the same size. That is, ${\cal V}_k \subset \{1,
\ldots, n\}$, $\bigcup_{k=1}^K {\cal V}_k = \{1, \ldots, n\}$, and ${\cal V}_k
\cap {\cal V}_{k'} = \emptyset$ for $k \neq k'$. For each $k$, its training
sample is ${\cal T}_k = \{1, \ldots, n\} \setminus {\cal V}_k$. Let $(g_{n,k},
\overline{Q}_{n,k,b})$ be the estimators of $(g_0, \overline{Q}_{0,b})$
constructed by fitting a data adaptive regression procedure using only data
available in the training sample ${\cal T}_k$. Then, letting $j(i)$ denote the
index of the validation set containing observation $i$, the empirical variance
of the cross-validated EIF is $\sigma^2_{n,\text{cv}, b} = \var
D_{n,\text{cv},b}$, where $D_{n,\text{cv},b}$ is the EIF evaluated at
$(g_{n,j(i)}, \overline{Q}_{n,j(i),b})$. The use of sample-splitting (i.e.,
cross-validation, cross-fitting) in constructing EIF-based estimators reduces
the need for theoretical regularity conditions and avoids overfitting of
nuisance estimators~\citep{bickel1993efficient, zheng2011cross}; we discuss any
advantages it may confer for variance estimation in subsequent sections.

\section{Semiparametric Variance Moderation}\label{methodology}

Application of TML estimation to construct targeted variable importance
estimates for a given set of biomarkers has been previously
considered~\citep{bembom2009biomarker}; however, marginal estimates of variable
importance are often insufficient or unreliable for deriving joint inference in
high-dimensional settings. Such approaches suffer significantly from instability
of standard error estimates in settings with limited sample sizes, erroneously
identifying differentially expressed biomarkers. This considerably limits their
utility in high-dimensional biomarker studies. In order to obtain stable joint
inference on a targeted variable importance measure across many biomarkers $b
= 1, \ldots, B$, we propose the use of variance moderation, which may be
achieved by applying the moderated t-statistic~\citep{smyth2004linear} to shrink
biomarker-specific estimates of sampling variability (based on the EIF) towards
a stabilized, pooled estimate.

As inference for $\psi_b$ is based on individual variability estimates
$\sigma_{n,b}$ (each derived from the corresponding EIF), our generalized
approach applies shrinkage directly to the estimated EIF $D_{n,b}$, yielding
a \textit{moderated EIF} $\tilde{D}_{n,b}$. The resultant moderated variance
estimate $\tilde{\sigma}^2_{n,b}$ is then the empirical variance of
$\tilde{D}_{n,b}$. The resultant stabilized variability estimates
$\tilde{\sigma}_{n,b}$ may directly be used in the construction of Wald-style
confidence intervals or the evaluation of hypothesis tests. Consider $B$
independent tests with null and alternative hypotheses $H_0: \psi_b = 0$ and
$H_1: \psi_b \neq 0$, and let $\psi_{n,b}$ denote either the one-step or TML
estimator of $\psi_b$; then, our proposal is as follows.
\begin{enumerate}
  \itemsep1pt
  \item Optionally, reduce the set of hypotheses by a filtering procedure, which
    may reduce the computational burden imposed by using flexible regression
    strategies for nuisance parameter estimation across many biomarker outcomes.
    As long as this initial filtering procedure does not affect the candidate
    biomarker rankings, its effect  may be readily accounted for in post-hoc
    multiple hypothesis testing corrections~\citep{tuglus2009modified}.
   \item For each biomarker,
    generate non/semi-parametric efficient estimates $\psi_{n,b}$ of $\psi_b$
    and corresponding estimates of the EIF $D_{n,b}(O_i)$, evaluated at the
    initial estimates of the nuisance parameters $(g_n, \overline{Q}_{n,b})$.
  \item Apply variance moderation across the biomarker-specific EIF estimates
    $(D_{n,b}: b = 1, \ldots, B)$ (e.g., via the \texttt{limma} \texttt{R}
    package~\citep{smyth2005limma}), constructing moderated variance estimates
    $\tilde{\sigma}_{n,b}^2$ for each biomarker. The moderated variance
    estimates are constructed by shrinking each  $\sigma_{n,b}^2$ towards the
    group variance across all other $(B-1)$ biomarkers.
    Equation~\eqref{moderated_t} gives the original
    formulation~\citep{smyth2004linear}; our procedure is analogous. Note that
    the variance moderation step is asymptotically inconsequential, that is,
    $\tilde{\sigma}_{n,b} \rightarrow \sigma_{n,b}$ as $n \rightarrow \infty$.
  \item For each biomarker-specific estimate of the target parameter
    $\psi_{n,b}$, construct a moderated t-statistic $(\tilde{t}_b: b = 1,
    \ldots, B)$ based on the corresponding moderated standard error estimate
    $\tilde{\sigma}_{n,b}$. The test statistic $\tilde{t}_b = \psi_{n,b}
    / \tilde{\sigma}_{n,b}$ may be used to evaluate evidence for the null
    hypothesis $H_0: \psi_b = 0$ of no treatment effect against the alternative
    $H_1: \psi_b \neq 0$. While the t-distribution with adjusted degrees of
    freedom~\citep{smyth2004linear} may be a suitable reference distribution for
    such test statistics, we advocate instead for use of a standardized logistic
    distribution (zero mean, unit variance). This alternative reference
    distribution exhibits subexponential tail behavior, allowing for
    conservative inference. In high-dimensional settings, the joint distribution
    of all $(\tilde{t}_b: b = 1, \ldots, B)$ test statistics may fail to
    converge quickly enough in $n$ to a $B$-dimensional multivariate normal or
    t-distribution, failing to control joint error appropriately. By contrast,
    the heavier tails of the logistic distribution provide more robust error
    control. Alternative approaches to conservative inference, e.g., via
    concentration inequalities~\citep{boucheron2013concentration} or Edgeworth
    expansions~\citep{gerlovina2017big}, may be suitable.
  \item Use a multiple testing correction to obtain accurate simultaneous
    inference across all $B$ biomarkers. A common approach is to use the
    Benjamini-Hochberg procedure to control the False Discovery
    Rate~\citep{benjamini1995controlling}, which controls Type-I error
    proportion in expectation in high-dimensional settings under conditions
    commonly considered acceptable in computational biology applications.
\end{enumerate}

Our proposed variance moderation procedure shrinks aberrant variability
estimates towards the center of their joint distribution, with a particularly
noticeable reduction of Type-I error when the sample size is small. Practically,
this approach limits the number of significant findings driven by unstable
estimates of the variance of $\psi_{n,b}$.

What's more, our proposal is convenient on account of its straightforward
application to the variance estimators based on the EIF and valid in all cases
where asymptotically linear estimators may be constructed. We stress that, since
our proposed procedure consists in a moderated variance estimator based on the
empirical variance of the estimated EIF, providing enhanced Type-I error rate
control is only guaranteed for multiple testing procedures that are based on
marginal hypothesis tests, as opposed to alternative techniques (e.g.,
permutation and resampling methods) that directly target the joint distribution
of test statistics~\citep{dudoit2008multiple}. To enhance accessibility, we have
made available an open source software implementation, the \texttt{biotmle}
package~\citep{hejazi2017biotmle, hejazi2020biotmlebioc}, available for the
\texttt{R} language and environment for statistical computing~\citep{R} through
the Bioconductor project~\citep{gentleman2004bioconductor} for computational
biology and bioinformatics.

\section{Simulation Studies}\label{simulation}

We evaluated our variance moderation strategy based on its Type-I error control
as assessed by the False Discovery Rate~\citep{benjamini1995controlling} (FDR).
We focus on the FDR owing to its pervasive use in addressing multiple hypothesis
testing in high-dimensional biology; however, our approach is equally compatible
with most post-hoc multiple testing corrections (e.g., Bonferroni's method to
control the family-wise error rate).
We assessed the relative performance of several data adaptive
non/semi-parametric estimators of the ATE, each using identical point estimation
methodology but different marginal variance estimators, and a single linear
modeling strategy in terms of their accuracy for joint inference.
We considered the performance of five variance estimation strategies: (1)
``standard'' variance moderation (via the \texttt{limma} \texttt{R}
package~\citep{smyth2005limma}) for the main-terms linear model; (2) a TML
estimator using the empirical variance of the full-sample EIF; (3) a TML
estimator using the empirical variance of the cross-validated EIF; (4) a TML
estimator with our variance moderation of the full-sample EIF; and (5) a TML
estimator with our variance moderation of the cross-validated EIF. For the
cross-validated variance estimators, we chose two-fold cross-validation based on
a conjecture that larger validation fold sizes would yield more conservative
variance estimates. We note that the one-step and TML estimators are
asymptotically equivalent and share a variance estimator, yet we use the TML
estimator on account of evidence of enhanced finite-sample
performance~\citep{vdl2011targeted}.
The TML estimators and their corresponding variance estimators were based on the
implementations in the \texttt{drtmle}~\citep{drtmlepackage} and
\texttt{biotmle}~\citep{hejazi2017biotmle, hejazi2020biotmlebioc} \texttt{R}
packages. To isolate the effect of variance moderation on FDR control, all
efficient estimator variants used the logistic reference distribution.

For these experiments, we simulated data from the following data-generating
mechanism. First, two baseline covariates are independently drawn as
$W_1 \sim \text{Uniform}(0, 1)$ and $W_2 \sim \text{Uniform}(0, 1)$. Next,
the exposure $A$ is drawn, conditionally on $\{W_1, W_2\}$, from
$A \mid W \sim \text{Bernoulli}[\text{expit}(0.5 + 2.5 W_1
- 3 W_2)]$. Finally, biomarker expression $Y_b$ is generated, conditionally on
$\{A, W_1, W_2\}$, by either $Y_{\text{null}} \mid A, W = 2 + W_1 + 0.5 W_2 +
W_1 \cdot W_2 + \epsilon_1$ or $Y_{\text{strong}} \mid A, W = 2 + W_1 +
0.5 W_2 + W_1 \cdot W_2 + 5 A + \epsilon_2$.
Throughout, $\text{expit}(x) = \{1 + \exp(-x)\}^{-1}$, $\epsilon_1 \sim
\text{Normal}(0, 1)$, and $\epsilon_2 \sim \text{Normal}(0, 0.2)$. The data on
a single observational unit are denoted by the random variable $O = (W_1, W_2,
A, (Y_b: 1, \dots, B))$, where each biomarker $(Y_b: 1, \dots, B)$ is generated
from $Y_{\text{strong}}$ or $Y_{\text{null}}$ depending on the setting. Note the
shared functional form of the outcome models, in particular that the interaction
term between $\{W_1, W_2\}$ gives rise to model misspecification issues when
linear regression is employed out-of-the-box. This design choice draws attention
to the advantages of relying upon non/semi-parametric efficient estimation
frameworks capable of incorporating data adaptive regression strategies (i.e.,
machine learning) in nuisance estimation.

For applications in which the exposure mechanism exhibits a lack of natural
experimentation (i.e., \textit{positivity violations}), estimation of the
exposure mechanism $g_n(A \mid W)$ can yield values extremely close to the
boundaries of the unit interval. Such extreme estimates compromise the
performance of data adaptive non/semi-parametric
estimators~\citep[e.g.,][]{moore2012causal}, in part due to the instability of
estimated inverse probability weights. Often, practical violations of the
positivity assumption occur when the exposure $A$ is strongly related to the
baseline covariates $W$, which manifests as an apparent lack of experimentation
of the exposure across covariate strata. To assess the impact of such violations
on variance estimation, we replace the exposure mechanism with $A \mid W \sim
\text{Bernoulli}\left(\text{expit}(0.5 + 2.5 W_1 - 3 W_2 - 2)\right)$ in a few
scenarios. Unlike the exposure mechanism above, which allows a minimum exposure
probability of $0.076$, this exposure mechanism allows a minimum exposure
probability of $0.011$, leading to positivity issues that may exacerbate bias
and variance instability in high dimensions.

To ensure compatibility of each of the efficient estimator variants, initial
estimates of the nuisance functions $g_n(A \mid W)$ and
$\overline{Q}_{n,b}(A,W)$ were constructed using the Super
Learner~\citep{vdl2007super} algorithm. The \texttt{SuperLearner} \texttt{R}
package~\citep{polley2019superlearner} was used to construct ensemble models
from a library of candidate algorithms that included linear or logistic
regression, regression with Bayesian priors, generalized additive
models~\citep{hastie2017generalized}, multivariate adaptive regression
splines~\citep{friedman1991multivariate}, extreme gradient boosted
trees~\citep{chen2016xgboost}, and random forests~\citep{breiman2001random}.

Here, we consider settings in which the exposure affects 10\% or 30\% of all
biomarkers. In each scenario, $B = 150$ biomarkers
are drawn from the equations for $Y_{\text{null}}$ and
$Y_{\text{strong}}$ in differing proportions. In any given simulation, we
consider observing $n$ i.i.d.~copies of $O$ for one of four sample sizes $n \in
\{50, 100, 200, 400\}$. Overall, we consider scenarios in which the number of
biomarkers exceeds the sample size as well as settings outside the
high-dimensional regime, i.e., $n/p = \{1/3, 2/3, 4/3, 8/3\}$. The former set of
scenarios emphasizes the utility of variance moderation when $p > n$, while
the latter demonstrates its negligible effect in larger samples.

Results are reported based on aggregation across $300$ Monte Carlo repetitions
for each scenario. In aggregate, these scenarios are used to evaluate the degree
to which each of the five variance estimation strategies controls the FDR.
Throughout, we restrict our attention to control of the FDR at the 5\% level, as
this is most commonly used in practice and the choice of threshold has no impact
on our proposed procedure. A few additional scenarios are considered in the
\href{sm}{Supplementary Materials}, including the relative estimator performance
in cases with no exposure effect and when there is a weaker exposure effect than
in the presently considered setting.

\begin{figure}[h!]
   \centering
   \includegraphics[scale=0.31]{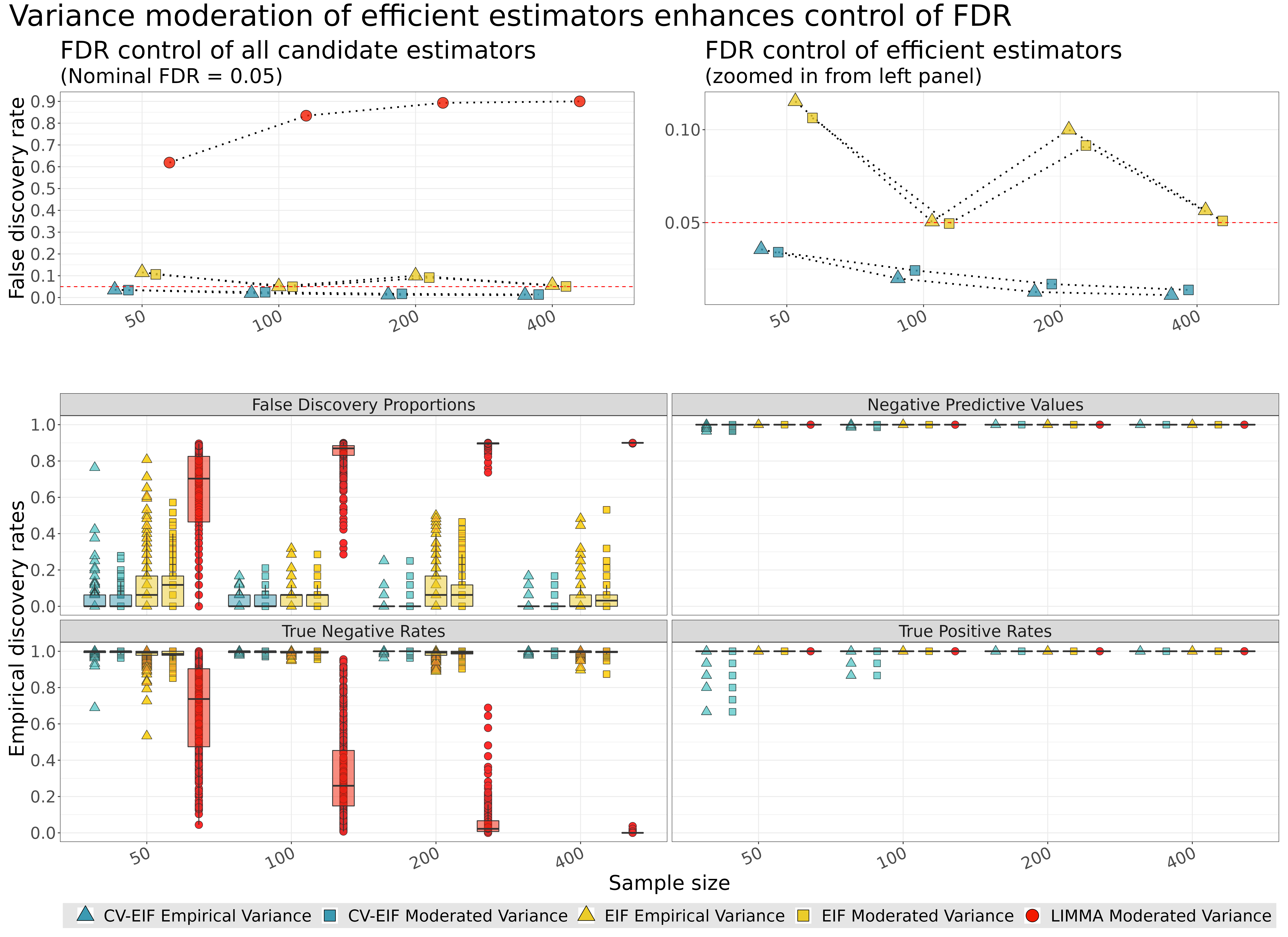}
   \caption{Control of the FDR across hypothesis testing procedures in a
     setting with strong exposure effect in 10\% of biomarkers and no
     positivity issues in the exposure mechanism. \textit{Upper panel}: Control
     of the FDR using the Benjamini-Hochberg correction. \textit{Lower panel}:
     Empirical distributions of false discovery proportions and negative
     predictive values, as well as of the true positive and true negative
     rates.}
  \label{fig:findings_min_goodpos}
\end{figure}

We begin with a scenario in which the effect of the exposure on biomarker
expression is strong, when the effect is either relatively rare (10\% of
biomarkers) or fairly common (30\% of biomarkers). In the rare effect setting,
expression values for the affected 10\% of biomarkers are generated by
$Y_{\text{strong}}$ while the values for the remaining 90\% arise from
$Y_{\text{null}}$. Here, we expect the efficient estimators with EIF-based
variance estimation strategies (whether moderated or not) to exhibit FDR control
approaching the nominal rate with increasing sample size while reliably
recovering truly differentially expressed biomarkers. Due to bias arising from
misspecification of the outcome model, the moderated linear model is expected to
perform poorly. The performance of the estimator variants is presented in
Figure~\ref{fig:findings_min_goodpos}.
As expected, variance-moderated hypothesis tests based on linear modeling fail
to control the FDR at the 5\% rate due primarily to model misspecification.
The efficient estimators based on the EIF exhibit reasonable performance, with
the full-sample variance estimators achieving the nominal rate by $n = 400$ and
the cross-validated variants consistently controlling the FDR more stringently
than the nominal rate. Examination of the false discovery proportions reveals
that variance moderation provides some benefit in improving FDR control at $n
= 50$, though this disappears quickly with increasing sample size. While the
true positive rates indicates good performance of all candidate procedures
(though the cross-validated variants are less reliable at smaller sample sizes),
the true negative rates demonstrate the consistent performance of the
cross-validated variants, performance improving with sample size for the
full-sample estimators, and degrading performance for the linear model.

\begin{figure}[h!]
   \centering
   \includegraphics[scale=0.31]{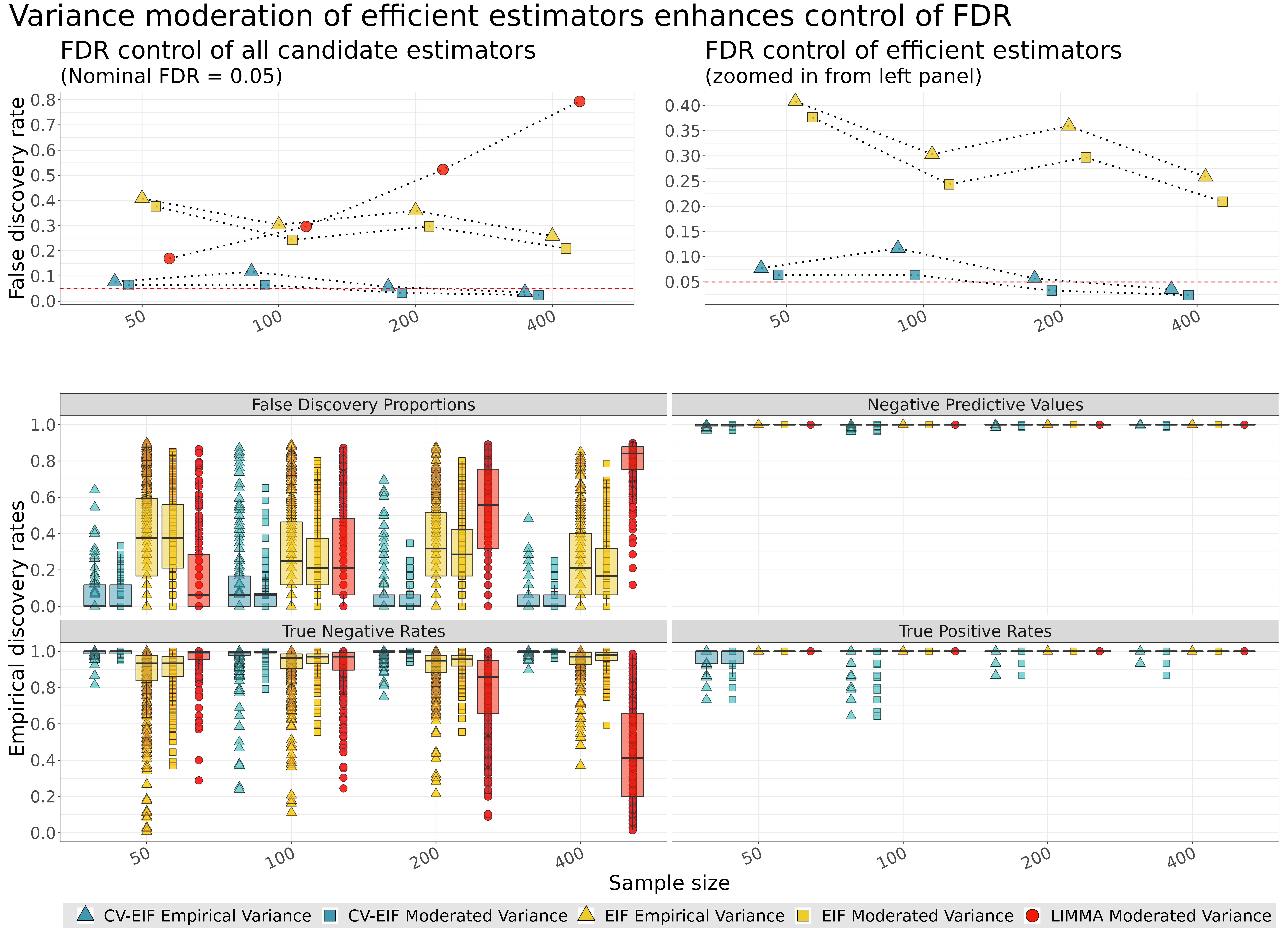}
  \caption{Control of the FDR across hypothesis testing procedures in a setting
    with strong exposure effect in 10\% of biomarkers and notable positivity
    issues in the exposure mechanism. \textit{Upper panel}: Control of the FDR
    using the Benjamini-Hochberg correction. \textit{Lower panel}: Empirical
    distributions of false discovery proportions and negative predictive
    values, as well as of the true positive and true negative rates.}
  \label{fig:findings_min_badpos}
\end{figure}

We now turn to a setting in which the exposure mechanism is prone to positivity
violations. In this case, the full-sample EIF-based variance estimators are
expected to exhibit relatively poor performance due to estimation instability in
the inverse probability weights; however, the cross-validated variants are
expected to provide FDR control at the nominal rate without sacrificing power.
Figure~\ref{fig:findings_min_badpos} presents the estimator performance.
As before, linear model-based hypothesis testing  fails to control the FDR at
the 5\% rate (owing to model misspecification). Positivity violations in the
exposure mechanism result in the full-sample EIF-based estimators yielding poor
FDR control as well. Their cross-validated counterparts fare significantly
better, achieving control at the nominal rate by $n = 200$. Both the FDR and
false discovery proportion panels illustrate that variance moderation of the
efficient estimators modestly but \textit{uniformly} improves their FDR control,
regardless of the use of sample-splitting in nuisance estimation. Consideration
of the true positive rates reveals good performance of all candidate procedures
(again, the cross-validated variants are slightly over-conservative). The true
negative rates show very strong control from the cross-validated variants and
worse but improving performance from the full-sample estimators; the linear
model displays unreliable, degrading performance. The protective effect of
variance moderation is made clear by the true negative rates.

\begin{figure}[h!]
  \centering
  \includegraphics[scale=0.31]{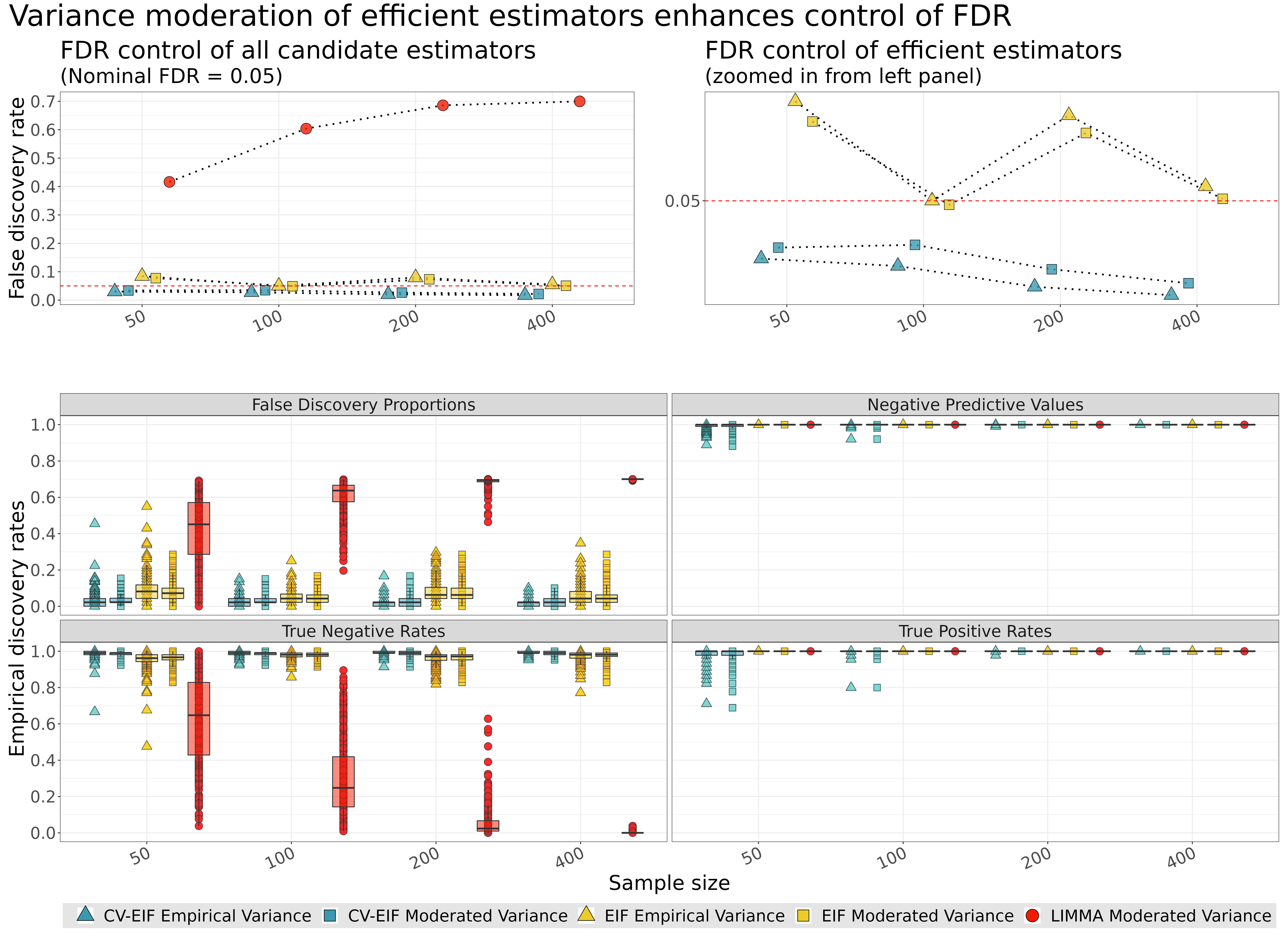}
  \caption{Control of the FDR across hypothesis testing procedures in a setting
    with strong exposure effect in 30\% of biomarkers and no positivity issues
    in the exposure mechanism. \textit{Upper panel}: Control of the FDR using
    the Benjamini-Hochberg correction. \textit{Lower panel}: Empirical
    distributions of false discovery proportions and negative predictive
    values, as well as of the true positive and true negative rates.}
  \label{fig:findings_mod_goodpos}
\end{figure}

Next, we turn to a setting in which the exposure has a strong effect on a larger
proprotion of biomarkers. This scenario is constructed by generating expression
values for 30\% of biomarkers from $Y_{\text{strong}}$ and the remaining 70\%
from $Y_{\text{null}}$. We begin with the exposure mechanism not prone to
positivity violations, in which case both the full-sample and cross-validated
efficient estimators are expected to exhibit FDR control near the nominal rate,
regardless of variance moderation. Due to model misspecification, the moderated
linear model is expected to exhibit poor FDR control.
Figure~\ref{fig:findings_mod_goodpos} visualizes the performance of the
candidate procedures.
Given that the exposure effect on biomarkers is more common, all of the
estimator variants fare comparatively better than in the rarer effect scenario
considered previously. As before, the poor performance of the linear modeling
strategy is caused by model misspecification bias. In comparison, the efficient
estimators all exhibit better performance, with the full-sample variance
estimators controlling the FDR at nearly the nominal rate and the
cross-validated variants providing more stringent control.
As with the prior setting summarized in Figure~\ref{fig:findings_min_goodpos},
the effect of variance moderation on FDR control is subtle, though examination
of the lower panel of Figure~\ref{fig:findings_mod_goodpos} reveals the stronger
error rate control that variance moderation achieves. While the true positive
rates reveal good performance from all candidate estimators by $n = 100$, the
true negative rates show slightly better control from the cross-validated
variants (relative to their full-sample counterparts); the linear model shows
poor performance at $n = 50$ and only degrades considerably thereafter.

\begin{figure}[h!]
   \centering
   \includegraphics[scale=0.31]{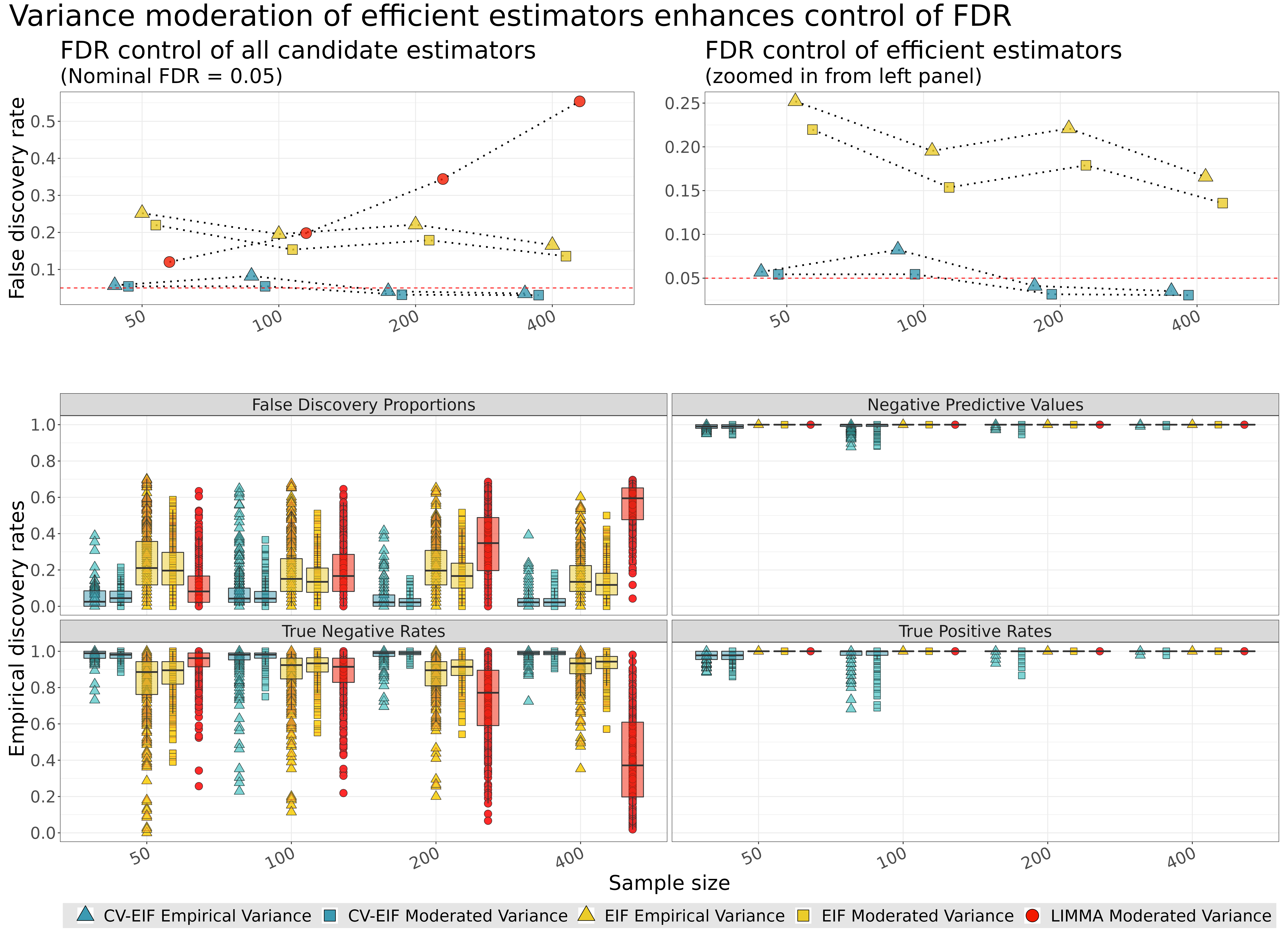}
   \caption{Control of the FDR across hypothesis testing procedures in a
     setting with strong exposure effect in 30\% of biomarkers and notable
     positivity issues in the exposure mechanism. \textit{Upper panel}: Control
     of the FDR using the Benjamini-Hochberg correction. \textit{Lower panel}:
     Empirical distributions of false discovery proportions and negative
     predictive values, as well as of the true positive and true negative
     rates.}
  \label{fig:findings_mod_badpos}
\end{figure}

Finally, we again consider an analogous setting in which the exposure mechanism
has positivity issues. As before, the linear modeling procedure is expected to
perform poorly. The efficient estimators with full-sample EIF-based variance
estimation ought to perform relatively poorly due to estimation instability
(from positivity violations) while the cross-validated variants are expected to
provide close-to-nominal FDR control. Figure~\ref{fig:findings_mod_badpos}
presents the results of exmaining the estimator variants in this setting.
The upper panel of Figure~\ref{fig:findings_mod_badpos} corroborates our
expectations about the linear modeling strategy's potential to yield erroneous
discoveries. While the linear model outperforms a subset of the efficient
estimators at $n = 50$, its performance degrades sharply thereafter. The
efficient estimators using full-sample EIF-based variance estimation display
relatively poor control of the FDR, failing to achieve the nominal rate but
maintaining their performance across sample sizes (unlike the linear model). The
estimator variants using cross-validated EIF-based variance estimation exhibit
far improved control of the FDR, nearly achieving the nominal rate in smaller
sample sizes and controlling the FDR more stringently in larger samples. A quick
examination of the lower panel of the figure makes clear the modest improvements
to error rate control that variance moderation provides. In particular, the true
positive rates are quite reliable for all candidate estimators, though the
cross-validated estimator variants are somewhat over-conservative in smaller
samples. By comparison, the true negative rates reveal the stronger control
that variance moderation confers for both the cross-validated and full-sample
estimator variants, and highlights the predictably poor performance of the
linear modeling strategy. Echoing results of the experiments presented in
Figure~\ref{fig:findings_min_badpos}, variance moderation improves FDR control
irrespective of whether sample-splitting is used.

While additional simulation studies and their results are presented in the
\href{sm}{Supplementary Materials}, our numerical investigations altogether
demonstrate the advantages conferred by applying variance moderation to
non/semi-parametric efficient estimators in settings with limited sample sizes
and a relatively large number of outcomes. In our experiments, the efficient
estimators have access to an eclectic library of machine learning algorithms for
nuisance estimation, significantly reducing the risk of model misspecification
bias. Generally, the full-sample EIF-based variance estimators exhibit poorer
FDR control than their cross-validated counterparts, suggesting a stabilizing
effect of sample-splitting on variance estimation, which itself pairs with
variance moderation. Our results reveal that variance moderation can have
substantial benefits in settings with positivity issues, which occur often in
observational studies. Overall, our findings suggest that variance moderation
can prove a useful and, at times, powerful tool for modestly improving FDR
control in high-dimensional settings, without adversely affecting the recovery
of truly differentially expressed biomarkers, and is especially useful in
high-dimensional settings when paired with cross-validation.

\section{Application in an Observational Smoking Exposure Study}\label{analysis}

We now apply our variance-moderated efficient estimation strategy to examine
evidence for differential methylation of CpG sites in whole blood as a result of
voluntary smoking exposure. Data for this illustrative application come from an
observational exposure study that enrolled 253 healthy volunteer participants
between 1993 and 1995 from the general population in Chapel Hill and Durham,
North Carolina. Among these participants, 172 self-reported as smokers and 81 as
nonsmokers (defined as having smoked fewer than 100 cigarettes in their
lifetime). For all participants, a limited set of baseline covariates, including
biological sex, race/ethnicity (minority status), and age, were recorded. The
study protocol and details on processing of biological samples have been
previously detailed~\citep{jones1993factors, bell1995occurrence,
su2016distinct}. DNA methylation levels of patients' whole blood DNA samples
were measured with the Infinium Human Methylation 450K BeadChip (Illumina,
Inc.), designed to measure methylation at $\approx$450,000 CpG sites across the
human genome. Prior analytic efforts~\citep{su2016distinct} normalized the raw
DNA methylation data via the ChAMP procedure~\citep{teschendorff2013beta,
morris2014champ} and deposited the processed $\beta$-values on the NCBI's Gene
Expression Omnibus (accession no.~GSE85210). In our re-analysis of this study,
we used these publicly available DNA methylation data, paired with phenotype
data provided by the study team.

For our differential methylation analysis, we used the aforementioned baseline
covariates as well as ``pack-years'' (self-reported packs of cigarettes
multiplied by years spent smoking) to adjust for potential baseline confounding
of the effect of smoking on DNA methylation. That DNA methylation varies
strongly across cell types has been well-studied and documented. Accordingly, we
followed standard practice in adjusting for cell-type composition of samples
from which DNA was collected by normalization against ``gold standard''
reference datasets~\citep{houseman2012dna, houseman2014reference}, accounting
for the relative abundance of CD4+ and CD8+ T-cells, natural killer cells,
B-cells, monocytes, and granulocytes. This form of adjustment disentangles the
effect of smoking on DNA methylation from the unwanted variation in DNA
methylation across cell types from which DNA samples were harvested. Our
differential methylation analysis strategy is summarized as follows.

First, the set of roughly 450,000 CpG sites was narrowed down by applying the
moderated linear modeling strategy of the \texttt{limma} \texttt{R}
package~\citep{smyth2005limma} to assess any association of differential
methylation with smoking, controlling for baseline covariates in the adjustment
set; the 2537 CpG sites with unadjusted p-values below the 5\% threshold were
advanced to the following stage. Next, using the \texttt{biotmle} \texttt{R}
package~\citep{hejazi2017biotmle,hejazi2020biotmlebioc}, our variance-moderated
non/semi-parametric efficient TML estimator was applied to evaluate evidence for
differential methylation attributable to smoking (based on the ATE), again
adjusting for the set of potential baseline confounders. Estimation of the
nuisance parameters $(g_n, \overline{Q}_{n,b})$ was performed using two-fold
cross-validation, and the Super Learner ensemble modeling
algorithm~\citep{vdl2007super, polley2019superlearner} was used to generate
out-of-sample predictions from a library of candidate algorithms that included
main-terms GLM regression, multivariate adaptive regression
splines~\citep{friedman1991multivariate}, and random
forests~\citep{breiman2001random}, among others.

Moderated test statistics were constructed to evaluate the null hypothesis of
no ATE at each CpG site, and testing multiplicity was accounted for by
adjusting the marginal p-values via Holm's procedure~\citep{holm1979simple},
thereby controlling the family-wise error rate (FWER). Marginal p-values for
each CpG site were generated by using the standardized normal distribution as
reference for the site-specific test statistics (the centered logistic
distribution proved too conservative when paired with the FWER metric);
moreover, Holm's procedure was chosen over alternative FWER-controlling
procedures as its rank-based nature satisfies previously outlined requirements
for error rate control in multi-stage analyses~\citep{tuglus2009modified}. Our
choice of FWER prioritizes conservative joint inference, complementing the more
lenient reference distribution and highlighting our proposal's flexibility. Our
analysis tagged 1173 CpG sites as differentially methylated by voluntary smoking
exposure.

The significantly differentially methylated CpG sites are located within the
\textit{AHRR}, \textit{ALPPL2/ALP1}, \textit{MYO1G}, \textit{F2RL3},
\textit{GFI1}, \textit{IER3}, \textit{HMHB1}, \textit{ITGAL}, \textit{LMO7},
\textit{GPR15}, \textit{NCOR2}, \textit{RARA}, \textit{SPOCK2}, \textit{HOX}
cluster, and \textit{RUNX3} genes, among others, agreeing with a prior analysis
of these data~\citep{su2016distinct}. Many of these genes have been linked to
disease ontology categories like hemotologic cancer, cardiovascular system
disease, hematopoietic system disease, and nervous system
cancer~\citep{su2016distinct}. In particular, the most significantly
differentially methylated CpG site, cg05575921, located in the \textit{AHRR}
gene, has been identified in over 30 epigenome-wide association studies on
smoking exposure in both blood and lung tissues~\citep{grieshober2020}.
Decreased methylation at this site is widely viewed as a robust biomarker of
smoking exposure~\citep{grieshober2020} and is associated with increased lung
cancer risk~\citep[e.g.,][]{fasanelli2015, zhang2016, bojesen2017, battram2019}.
Table~\ref{supp:tab:anno_top50}, in the \href{sm}{Supplementary Materials},
presents the top 50 differentially methylated CpG sites.

\begin{figure}[h!]
   \centering
   \includegraphics[scale=0.34]{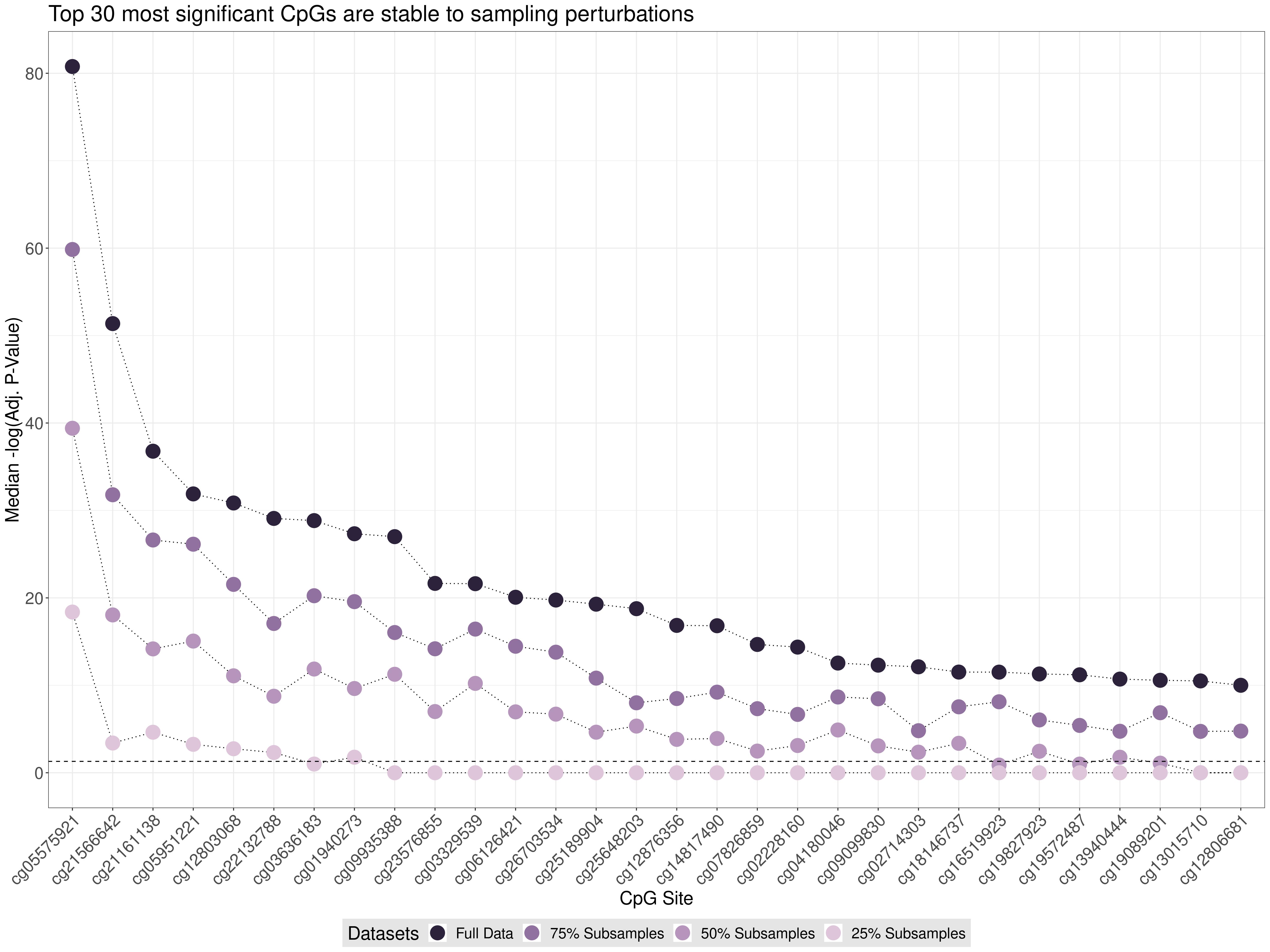}
  \caption{Evaluation of the top 30 differentially methylated CpGs (orderd left
    to right) from the complete analysis in terms of median
    \{$-\log_{10}(\text{adj.~p-value})$\}'s across the three subsampling
    schemes.}
  \label{fig:cpg_ranks}
\end{figure}

Despite the close agreement between the top set of differentially methylated
CpGs revealed by our analysis and those identified in prior analyses, we
questioned the stability of our proposal for real-world data analysis. To assess
this, we designed and conducted an empirical sensitivity analysis that
subsampled study units to capture the effect of data removal on the ranking of
differentially methylated CpG sites. The procedure was carried out by sampling
without replacement $\{25\%, 50\%, 75\%\}$ of study units, performing our
proposed analysis (as described above) to generate a ranked list of CpG sites,
and comparing these top CpG sites against those identified in the complete-data
analysis. Since the sensitivity of the preliminary filtering step to subsampling
does not relate directly to our procedure's stability, we restricted each of
these analyses only to the 2537 CpG sites that passed the filtering step of the
complete-data analysis. For each subsampling proportion, this sensitivity
analysis strategy was repeated 10 times, allowing for the frequency with which
CpGs were tagged as differentially methylated to be evaluated.
Figure~\ref{fig:cpg_ranks} displays the results of our sensitivity analysis.
Cursory examination of Figure~\ref{fig:cpg_ranks} reveals that our findings
concerning the top 30 differentially methylated CpG sites are robust to a loss
of 25\% of study units, as the median adjusted p-values of all of these CpG
sites exceed the 5\% detection threshold at the 75\% subsampling level. Upon
further reductions in sample size, the differential methylation signal is still
fairly reliable: the median adjusted p-values for $\approx$75\% of the CpG sites
(the top 23) exceed the detection threshold even when 50\% of study units have
been removed. Finally, this form of evidence for differential methylation shows
that the top 6 CpG sites identified by our analysis are robust to a loss of as
much as 75\% of the data, meaning that these same CpGs could have been tagged as
differentially methylated had the study included as few as 64 units (instead of
the 253 units actually enrolled). Note that while the adjusted p-values reported
for each of the 30 CpGs in the figure are the medians across the 10 iterations
for each of the subsampling schemes, those for the complete-data analysis are
not medians (i.e., that analysis was only run once).
Figure~\ref{supp:fig:parallel_coord} in the \href{sm}{Supplementary Materials}
presents an extension of Figure~\ref{fig:cpg_ranks}, showing how the minimum,
median, and maximum adjusted p-values vary across subsampling schemes for the
top 30 differentially methylated CpGs. Altogether, this sensitivity analysis
demonstrates that our differential methylation procedure reliably recovers
evidence for biologically meaningful findings, with power only beginning to
degrade significantly with major losses in sample size.



\section{Discussion}\label{discussion}

We have proposed a novel procedure for stabilizing non/semi-parametric efficient
estimators of scientifically relevant statistical parameters, combining distinct
lines of inquiry on variance moderation and sample-splitting principles in the
process. Our variance moderation procedure may be applied directly to the
standard variance estimator of regular and asymptotically linear estimators in
the nonparametric model, i.e., the efficient influence function. These efficient
estimators are capable of incorporating machine learning in nuisance estimation,
curbing the risk of model misspecification bias, which limits the reliability of
parametric modeling approaches. Our variance moderation technique improves the
inferential stability of hypothesis testing based on these efficient estimators
in high-dimensional settings, and, when combined with cross-validation, it is
capable of providing reliably conservative joint inference. Our proposal amounts
to an automated procedure for using these state-of-the-art estimators to obtain
valid joint inference in high-dimensional biomarker studies while circumventing
the pitfalls of model misspecification bias, sampling distribution instability,
and anti-conservative variance estimation.

Our demonstration of this proposal focused on efficient estimators of the
average treatment effect; however, the outlined procedure can be readily adapted
to any regular and asymptotically linear estimator, accommodating extensions to
a wide variety of parameters of scientific interest. Notable areas for future
adaptation of this methodology include recently developed estimators of the
causal effects of continuous exposures~\citep{diaz2012population,
hejazi2020efficient} and those of causal mediation effects tailored for path
analysis~\citep{diaz2020causal, hejazi2021nonparametric}. Our simulation
experiments highlight the benefits conferred by our strategy, both in
conjunction with and in the absence of sample-splitting, showing that variance
moderation can modestly but uniformly improve Type-I error control in several
common scenarios. In a secondary re-analysis of DNA methylation data from an
observational study on the epigenetic effects of smoking, we show our procedure
to be capable of recovering differentially methylated CpG sites identified in
prior analyses and validated in biological experiments; moreover, a sensitivity
analysis reveals the findings of our approach to be highly stable even with
artificially diminished sample sizes. Given the utility of the procedure, we
have developed the free and open source \texttt{biotmle} \texttt{R}
package~\citep{hejazi2017biotmle, hejazi2020biotmlebioc} and contributed it to
the Bioconductor project~\citep{gentleman2004bioconductor}, making this novel
strategy easily accessible to the computational biology scientific community.

\section*{Acknowledgments}
We thank M.T.~Smith, N.~Rothman, and Q.~Lan for helpful discussions about an
alternative real-world data analysis example. We thank D.~Bell for providing the
data used in the real-world data analysis and for helpful correspondence on the
study details for the application presented. We are grateful to S.~Dudoit for
numerous helpful discussions on data visualization and the presentation of
results.

\section*{Funding}
The author(s) disclosed receipt of the following financial support for the
research, authorship, and/or publication of this article: NSH's work was
supported in part by the National Institute of Environmental Health Sciences
[award no.~R01-ES021369] and the National Science Foundation [award no.~DMS
2102840]. M.T.~Smith, N.~Rothman, and Q.~Lan also received support from the
National Institute of Environmental Health Sciences [award no.~P42-ES004705] and
the National Cancer Institute.

\bibliography{refs}
\end{document}


\label{supplement_simulation}
\maketitle

\section{Results of Additional Simulation Studies}\label{more_sims}

We report the results of several additional simulation experiments. Firstly, we
consider a scenario in which there is no effect of the exposure at all (i.e.,
the ``global null''). Secondly, we examine a setting in which the effect of the
exposure is attenuated relative to the scenario considered in the
\href{paper}{main manuscript}.
Unless otherwise stated, all other aspects of the data-generating process and
simulation study design remain as previously described.

\subsection{Simulation \#2: Global null of no effect of
  exposure}\label{sim_no_effect}

Here, we examine the performance of the candidate estimators in term of their
control of the FDR under a global null hypothesis of no effect of exposure on
any biomarkers. In this setting, expression values for all biomarkers are
generated by $Y_{\text{null}}$. We note that such concepts as ``statistical
power'' and ``true positive rate'' are undefined in the absence of any truly
differentially expressed biomarkers. We again examine two cases --- one in which
the exposure mechanism does not allow practical violations of the assumption of
positivity and another in which it is prone to such violations.

\begin{figure}[H]
  \centering
  \includegraphics[origin=c,scale=0.3]{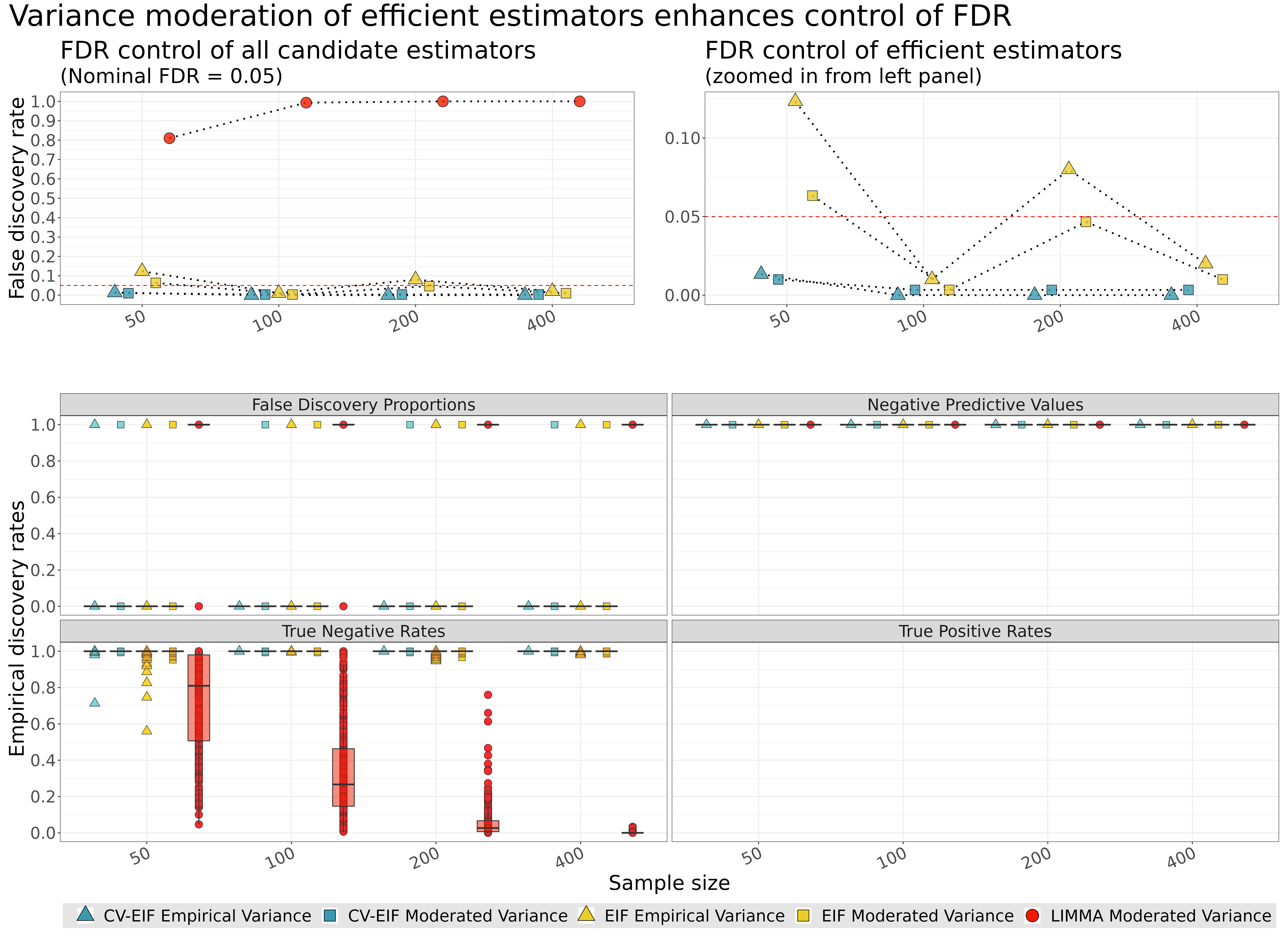}
  \caption{Control of the FDR across hypothesis testing procedures in a
    setting with no effect of exposure on biomarkers and no significant
    positivity issues in the exposure mechanism. \textit{Upper panel}: Control
    of the FDR using the Benjamini-Hochberg correction. \textit{Lower panel}:
    Empirical distributions of false discovery proportions and negative
    predictive values, as well as of the true positive and true negative
    rates.}
  \label{fig:global_null_goodpos}
\end{figure}

When the exposure mechanism allows sufficient natural experimentation (i.e., no
significant positivity violations), the efficient estimators are expected to
perform well, as these estimators do not suffer from instability introduced by
extreme inverse probability weights. As such, the efficient estimators are
expected to exhibit FDR control nearing the nominal 5\% rate with increasing
sample size. Further, we expect variance moderation to have little notable
effect on estimator performance. The linear modeling procedure is expected to
perform poorly due to model misspecification bias. The performance of the
estimator variants is displayed in Figure~\ref{fig:global_null_goodpos}.
As expected, the moderated linear model performs poorly, tagging over 75\% of
biomarkers as differentially expressed at $n = 50$ and quickly proceeding to
incorrectly label all biomarkers as such in larger samples. Variance moderation
improves FDR control of the full-sample efficient estimators to nearly the
nominal rate at $n = 50$, with more stringent control in larger sample sizes. By
contrast, without variance moderation, the same variance estimator fails to
control the FDR at the nominal rate at $n = 50$ and $n = 200$. Here, the benefit
of applying variance moderation to the full-sample efficient estimators is
clear. The cross-validated efficient estimators are more conservative than their
full-sample counterparts, as expected. Unsurprisingly, variance moderation
appears to benefit these estimators less so than it does their full-sample
counterparts. Importantly, these cross-validated efficient estimators provide
more stringent FDR control than the allotted 5\% (uniformly across the sample
sizes considered), a reassuring finding given that the true positive rate is 0\%
in this setting.

\begin{figure}[H]
  \centering
  \includegraphics[origin=c,scale=0.3]{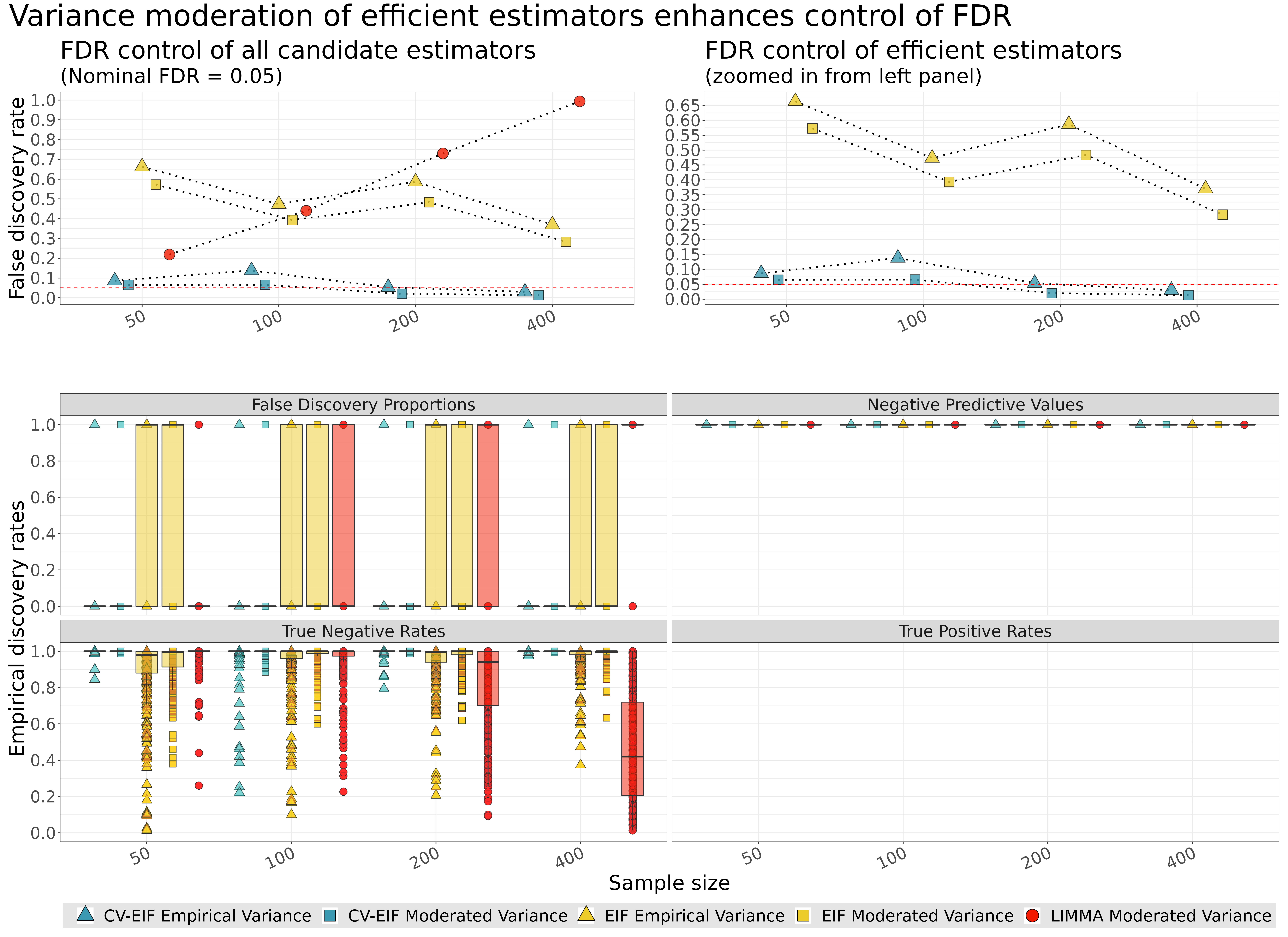}
  \caption{Control of the FDR across hypothesis testing procedures in a
    setting with no effect of exposure on biomarkers and severe positivity
    issues in the exposure mechanism. \textit{Upper panel}: Control of the FDR
     using the Benjamini-Hochberg correction. \textit{Lower panel}: Empirical
     distributions of false discovery proportions and negative predictive
     values, as well as of the true positive and true negative rates.}
  \label{fig:global_null_badpos}
\end{figure}

When the exposure mechanism yields positivity violations, the full-sample
efficient estimators are expected to perform poorly in controlling the FDR
(again due to instability of estimated inverse probability weights) while their
cross-validated counterparts should provide more stringent control. Variance
moderation is expected to improve the performance of both classes of estimators,
as pooling variance estimates can allow the relatively rare deviations arising
from positivity violations to be ``smoothed out'' across biomarkers. As in all
other cases, the moderated linear model is expected to perform poorly due to
bias from misspecification of the outcome model.
Figure~\ref{fig:global_null_badpos} presents the results of examining the
estimator variants in this setting.
Inspection of the figure reveals that the moderated linear model provides better
control of the FDR than the full-sample efficient estimators at the lowest
sample size ($n = 50$) but proceeds to incorrectly tag very high proportions of
biomarkers as being differentially expressed as sample size grows. The
full-sample efficient estimators provide poor control of the FDR, on account of
positivity violations. In particular, these estimators fail to control the FDR
at a reasonable rate uniformly across the considered sample sizes. On the other
hand, the cross-validated efficient estimators provide much better FDR control.
Across both classes of these estimators, variance moderation improves FDR
control, highlighting the benefits of using our proposed variance moderation
technique with EIF-based variance estimation. Importantly, variance moderation
of the cross-validated estimators allows the FDR to be controlled at nearly the
nominal rate irrespective of sample size, demonstrating that even conservative
variance estimation based on sample-splitting principles stands to benefit from
this form of variance moderation.

\subsection{Simulation \#3: Weak effect of exposure}\label{sim_weak_effect}

\begin{figure}[H]
  \centering
  \includegraphics[origin=c,scale=0.3]{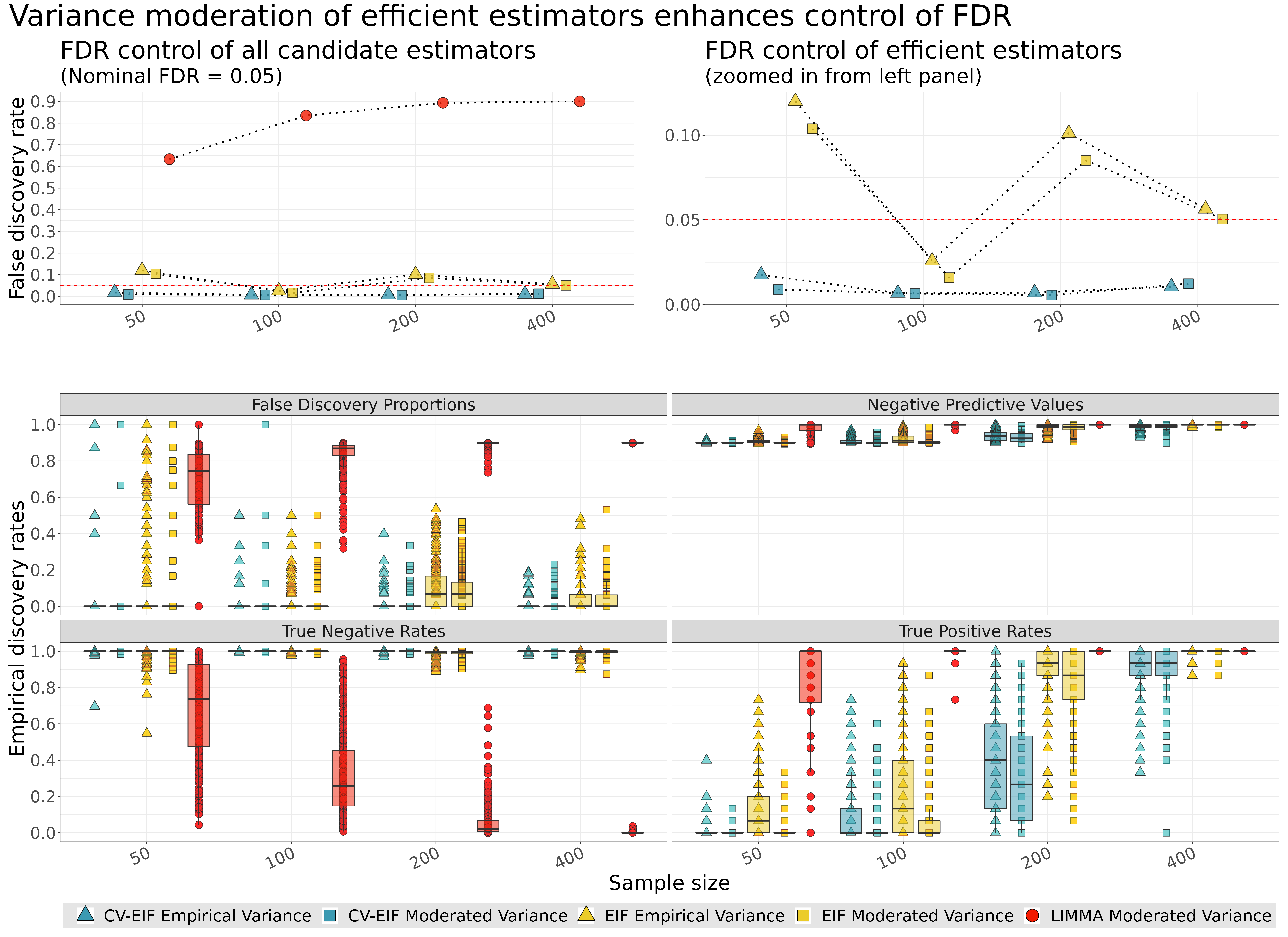}
  \caption{Control of the FDR across hypothesis testing procedures in a setting
    with weak effect of exposure on 10\% of biomarkers and no significant
    positivity issues in the exposure mechanism. \textit{Upper panel}: Control
    of the FDR using the Benjamini-Hochberg correction. \textit{Lower panel}:
    Empirical distributions of false discovery proportions and negative
    predictive values, as well as of the true positive and true negative rates.}
  \label{fig:findings_min_weak_effect}
\end{figure}

In another set of numerical experiments, we examine the performance of the
candidate estimators in controlling the FDR in the absence of a strong
exposure effect, using instead an exposure effect 97\% smaller than that
appearing in the alternative scenario in the \href{paper}{main manuscript}. To
achieve this effect, we replace the structural equation for $Y_{\text{strong}}$
with one in which the effect of the exposure is attenuated: $Y_{\text{weak}}
\mid A, W = 2 + W_1 + 0.5 W_2 + W_1 \cdot W_2 + 0.15 A + \epsilon$, where
$\epsilon \sim \text{Normal}(0, 0.2)$. As before, we consider how each procedure
fares when the effect of the exposure is rare (10\% of biomarkers) and relatively
common (30\% of biomarkers). We begin by examining the rare effect setting, in which
the expression values for 10\% of biomarkers were generated by $Y_{\text{weak}}$
while values for the remaining 90\% were assigned by $Y_{\text{null}}$. In this
case, we expect the efficient (EIF-based) estimators to control the to control
the FDR at the nominal rate while reliably recovering truly differentially
expressed biomarkers at larger sample sizes, regardless of the use of variance
moderation. This is due to the relatively weak effect, which makes accurately
tagging biomarkers as differentially expressed more challenging. Due to model
misspecification bias, the linear modeling procedure is expected to perform
poorly across all scenarios. The relative performance of the estimators is
presented in Figure~\ref{fig:findings_min_weak_effect}.
As expected, the linear model fails to control the FDR at the nominal rate of
5\%, while the efficient estimation procedures exhibit far better performance.
These estimators achieve FDR control at the nominal rate at $n = 200$ and $n
= 400$ without sample-splitting and provide more stringent control with
sample-splitting. The upper panel of the figure makes clear that the efficient
estimators ought to be preferred over parametric modeling procedures; moreover,
in the full-sample case, there is a modest but noticeable effect of variance
moderation on control of the FDR. This effect is even more clearly visible in
the lower panel depicting the false discovery proportion. Here, it is clear that
variance moderation can improve the small-sample performance of these
estimators. The cross-validated variance estimators appear extremely
conservative, controlling the FDR at a rate well below the nominal 5\% level,
which may prove problematic in a setting with a weak effect present in only
a relatively small set of biomarkers.

\begin{figure}[H]
  \centering
  \includegraphics[origin=c,scale=0.3]{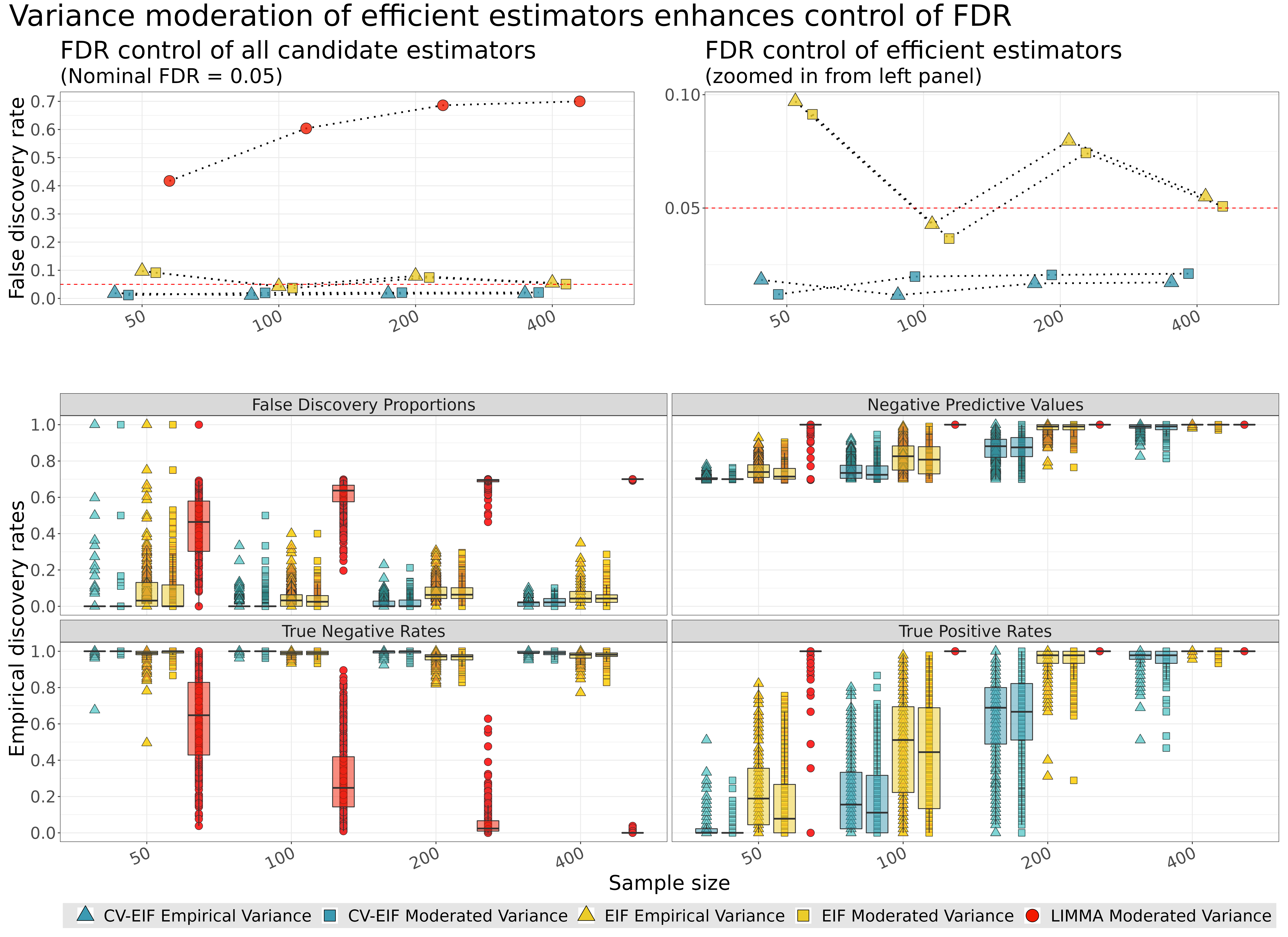}
  \caption{Control of the FDR across hypothesis testing procedures in a setting
    with weak effect of exposure on 30\% of biomarkers and no significant
    positivity issues in the exposure mechanism. \textit{Upper panel}: Control
    of the FDR using the Benjamini-Hochberg correction. \textit{Lower panel}:
    Empirical distributions of false discovery proportions and negative
    predictive values, as well as of the true positive and true negative rates.}
  \label{fig:findings_mod_weak_effect}
\end{figure}

Next, we consider a setting in which the number of biomarkers affected by the
exposure is larger, with 30\% of biomarker outcomes being generated by
$Y_{\text{weak}}$ and the remaining 70\% by $Y_{\text{null}}$.
Figure~\ref{fig:findings_mod_weak_effect} visualizes the relative performance of
the candidate procedures. As before, we expect the efficient estimation
procedures to control the FDR at close to the nominal rate in larger samples,
with more conservative control provided by the cross-validated variants. The
linear modeling strategy is expected to perform poorly due to model
misspecification bias.
Examination of Figure~\ref{fig:findings_mod_weak_effect} reveals that the
efficient estimators provide FDR control close to the nominal rate when
sample-splitting is not used in estimation of the nuisance parameters, though
the degree of control varies somewhat with sample size. The cross-validated
variants of these estimators uniformly provide FDR control more conservative
than the nominal rate, risking failure to identify a subset of true findings. In
spite of this, inspection of the lower panel of the figure reveals that all of
the efficient estimators achieve good true negative and true positive rates on
average (across simulations). Throughout, the effect of variance moderation on
the efficient estimators is quite small but clearly visible in inspecting the
metrics displayed in the lower panel of the figure. As expected, model
misspecification bias compromises the performance of the linear modeling
procedure.

\section{Annotated Results of Differential Methylation
Analysis}\label{anno_results}

Table~\ref{tab:anno_top50} presents the annotated results for the top 50 CpG
sites identified by our nonparametric differential methylation analysis as
described in the \href{paper}{main manuscript}. As stated previously, these
results agree quite strongly with those of the original analysis of this study's
dataset~\citep{su2016distinct}, which relied upon a parametric modeling-based
analytic strategy that imposed more restrictive modeling assumptions. As our
analysis tagged 1173 CpG sites as differentially methylated, we refrain from
presenting the complete set of results here, in the interest of readability. The
complete set of results are available in a publicly accessible GitHub repository
(TODO: URL HERE), which serves as a complete collection of \texttt{R} scripts
required to reproduce our analysis.
\begin{table}

\caption{\label{tab:anno_top50}Top 50 CpG sites (ranked by adjusted p-value) tagged by our nonparametric differential methlation analysis.}
\centering
\begin{tabular}[t]{llrlll}
\toprule
CpG & Chromosome & Position & Adj. P-value & Gene Name (UCSC) & CpG Island Relation\\
\midrule
cg05575921 & chr5 & 373378 & 1.69e-81 & AHRR & N\_Shore\\
cg21566642 & chr2 & 233284661 & 4.26e-52 & NA & Island\\
cg21161138 & chr5 & 399360 & 1.68e-37 & AHRR & OpenSea\\
cg05951221 & chr2 & 233284402 & 1.28e-32 & NA & Island\\
cg12803068 & chr7 & 45002919 & 1.4e-31 & MYO1G & S\_Shore\\
\addlinespace
cg22132788 & chr7 & 45002486 & 8.18e-30 & MYO1G & Island\\
cg03636183 & chr19 & 17000585 & 1.45e-29 & F2RL3 & N\_Shore\\
cg01940273 & chr2 & 233284934 & 4.63e-28 & NA & Island\\
cg09935388 & chr1 & 92947588 & 9.95e-28 & GFI1;GFI1;GFI1 & Island\\
cg23576855 & chr5 & 373299 & 2.21e-22 & AHRR & N\_Shore\\
\addlinespace
cg03329539 & chr2 & 233283329 & 2.39e-22 & NA & N\_Shore\\
cg06126421 & chr6 & 30720080 & 8.46e-21 & NA & OpenSea\\
cg26703534 & chr5 & 377358 & 1.77e-20 & AHRR & S\_Shelf\\
cg25189904 & chr1 & 68299493 & 5.25e-20 & GNG12 & S\_Shore\\
cg25648203 & chr5 & 395444 & 1.72e-19 & AHRR & OpenSea\\
\addlinespace
cg12876356 & chr1 & 92946825 & 1.4e-17 & GFI1;GFI1;GFI1 & Island\\
cg14817490 & chr5 & 392920 & 1.53e-17 & AHRR & OpenSea\\
cg07826859 & chr7 & 45020086 & 2.12e-15 & MYO1G & OpenSea\\
cg02228160 & chr5 & 143192067 & 4.25e-15 & HMHB1 & OpenSea\\
cg04180046 & chr7 & 45002736 & 2.85e-13 & MYO1G & Island\\
\addlinespace
cg09099830 & chr16 & 30485485 & 4.94e-13 & ITGAL;ITGAL & Island\\
cg02714303 & chr13 & 76334728 & 7.66e-13 & LMO7;LMO7 & OpenSea\\
cg18146737 & chr1 & 92946700 & 3.06e-12 & GFI1;GFI1;GFI1 & Island\\
cg16519923 & chr16 & 30485810 & 3.1e-12 & ITGAL;ITGAL & S\_Shore\\
cg19827923 & chr2 & 231790777 & 5.03e-12 & GPR55 & OpenSea\\
\addlinespace
cg19572487 & chr17 & 38476024 & 6.26e-12 & RARA;RARA;RARA & S\_Shore\\
cg13940444 & chr12 & 53617382 & 1.94e-11 & RARG & S\_Shelf\\
cg19089201 & chr7 & 45002287 & 2.67e-11 & MYO1G & Island\\
cg13015710 & chr12 & 125039343 & 3.14e-11 & NA & Island\\
cg12806681 & chr5 & 368394 & 9.91e-11 & AHRR & N\_Shore\\
\addlinespace
cg14675361 & chr13 & 76334583 & 1.09e-10 & LMO7;LMO7 & OpenSea\\
cg16391678 & chr16 & 30485597 & 1.17e-10 & ITGAL;ITGAL & Island\\
cg04982781 & chr22 & 39714193 & 1.36e-10 & RPL3;RPL3;RNU86 & N\_Shore\\
cg19859270 & chr3 & 98251294 & 1.63e-10 & GPR15 & OpenSea\\
cg11902777 & chr5 & 368843 & 2.19e-10 & AHRR & N\_Shore\\
\addlinespace
cg00931843 & chr6 & 155442993 & 1.14e-09 & TIAM2 & OpenSea\\
cg21611682 & chr11 & 68138269 & 1.23e-09 & LRP5 & OpenSea\\
cg23161492 & chr15 & 90357202 & 1.38e-09 & ANPEP & N\_Shore\\
cg03450842 & chr10 & 80834947 & 1.46e-09 & ZMIZ1 & OpenSea\\
cg08709672 & chr1 & 206224334 & 1.5e-09 & AVPR1B;AVPR1B & S\_Shore\\
\addlinespace
cg14781374 & chr10 & 101998405 & 1.85e-09 & CWF19L1;SNORA12 & OpenSea\\
cg15342087 & chr6 & 30720209 & 3.15e-09 & NA & OpenSea\\
cg07178945 & chr12 & 4488800 & 5.7e-09 & FGF23;FGF23 & OpenSea\\
cg01440841 & chr4 & 154681066 & 6.61e-09 & RNF175;RNF175 & Island\\
cg10691866 & chr7 & 65817282 & 7.4e-09 & TPST1 & OpenSea\\
\addlinespace
cg04885881 & chr1 & 11123118 & 7.55e-09 & NA & S\_Shelf\\
cg17507897 & chr20 & 17943694 & 7.71e-09 & SNX5;SNORD17;SNX5 & OpenSea\\
cg11183632 & chr20 & 21503152 & 9.23e-09 & NA & Island\\
cg12459932 & chr1 & 25292018 & 1.6e-08 & RUNX3 & OpenSea\\
cg08354053 & chr17 & 30630872 & 2.03e-08 & RHBDL3 & OpenSea\\
\bottomrule
\end{tabular}
\end{table}

\section{Extended Stability Analysis of Differential Methylation
Results}\label{stable_ext}

\begin{figure}[h!]
   \centering
   \includegraphics[scale=0.34]{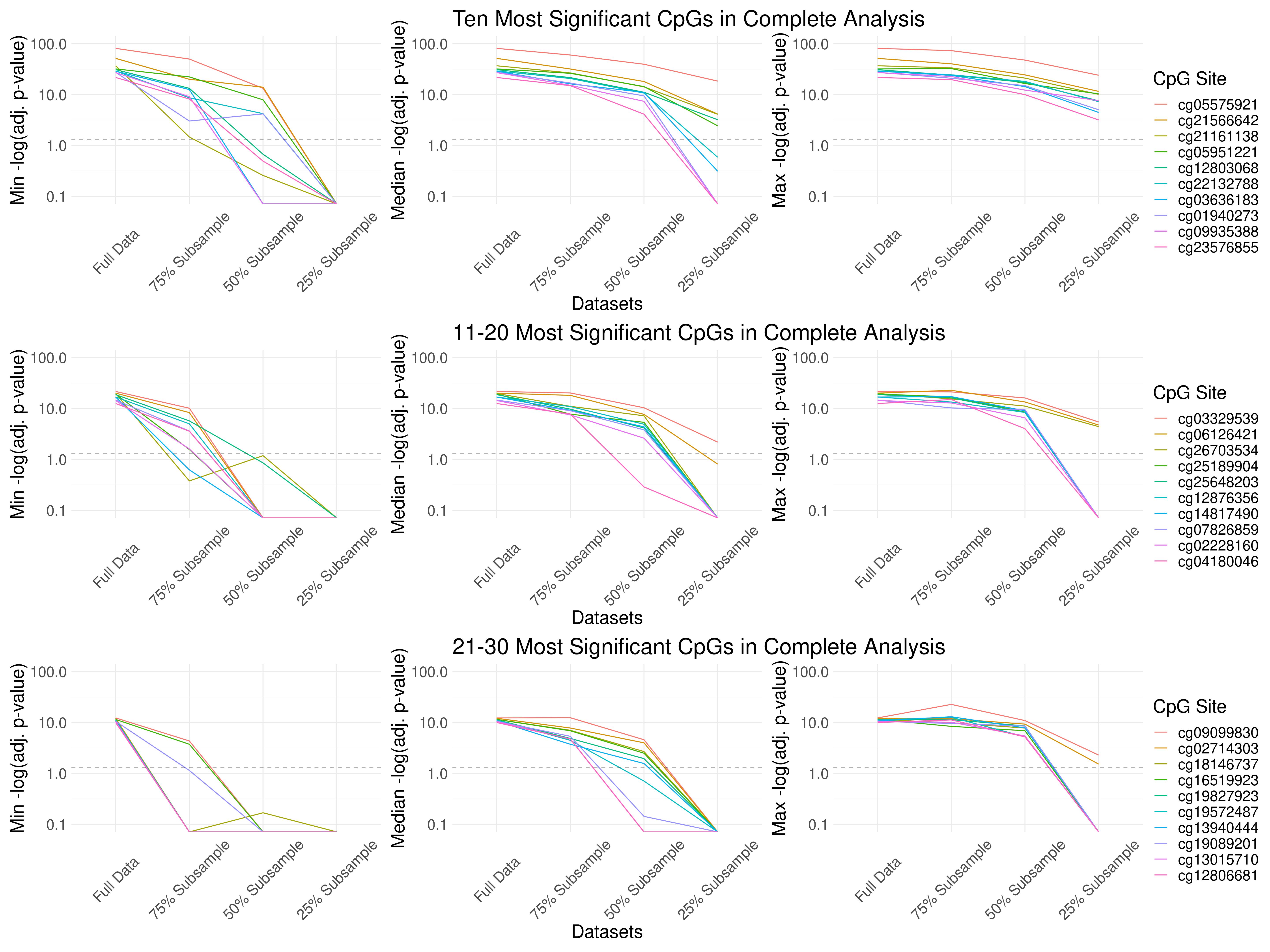}
   \caption{Evaluation of the top 30 differentially methylated CpGs from the
     complete analysis in terms of minimum, median, and maximum
     \{$-\log_{10}(\text{adj.~p-value})$\}'s across the three subsampling
     proportions.}
  \label{fig:parallel_coord}
\end{figure}

Figure~\ref{fig:parallel_coord} provides an extended view of the stability
analysis of our differential methylation results presented in the
\href{paper}{main manuscript}. Each panel of Figure~\ref{fig:parallel_coord}
diplays how the minimum, median, or maximum of the adjusted p-values (across
iterations) for a subset of the top 30 differentially methylated CpG sites vary
with decreases in sample size. Figure~\ref{fig:parallel_coord} is complementary
to Figure~\ref{main:fig:cpg_ranks} in that the central panel of the former
corresponds exactly to the results presented in the latter.
While each row of Figure~\ref{fig:parallel_coord} focuses on the top 10,
11\textsuperscript{th}--20\textsuperscript{th}, or
21\textsuperscript{st}--30\textsuperscript{th} CpG sites, each column
highlights one of the adjusted p-value metrics (i.e., minimum, median, or
maximum across the 10 analysis iterations). Examination of the figure reveals
that the strongest evidence for differential methylation (in the rightmost
column) does not appreciably wane for the top 10 CpGs and only decreases for
a limited subset of the 11\textsuperscript{th}--30\textsuperscript{th} CpGs once
75\% of the study units have been removed. The central column indicates that the
median adjusted p-values are fairly insensitive for at least the top 20 CpG
sites, as 50\% of the study units must be removed before the magnitude of these
adjusted p-values falls below the 5\% detection threshold. Inspection of the
leftmost column shows that, even in the worst case, the strength of evidence for
differential methylation is stable for the top 10 CpGs when 25\% of the study
units are removed.

\bibliography{refs}